\documentclass[fleqn,12pt]{wlscirep}

\usepackage[utf8]{inputenc}
\usepackage[T1]{fontenc}

\usepackage{lineno}

\usepackage{verbatim}
 \usepackage{setspace}
\usepackage{soul}
\soulregister\cite7
\soulregister\ref7

\usepackage{seqsplit}
\usepackage{gensymb}
\usepackage{dirtytalk}
\usepackage{outlines}
\usepackage{esdiff}
\usepackage{csquotes}

\usepackage{amsmath}

\newcommand{\Themap}{Signal propagation map}
\newcommand{\themap}{signal propagation map}

\newcommand{\sfigspotwavelength}{\ref{sfig:spot_wavelength}}
\newcommand{\sfigbleaching}{\ref{sfig:bleaching}}

\newcommand{\sfigDFonly}{\ref{sfig:DF_only}}

\newcommand{\sfigkernels}{\ref{sfig:kernels}}

\newcommand{\sfiguncvsWTcontrol}{\ref{sfig:unc31_vs_WT_control}}
\newcommand{\sfiguncvsWTstats}{\ref{sfig:unc31_vs_WT_stats}}

\newcommand{\sfigRIDmoreresponses}{\ref{sfig:RID_more_responses}}

\newcommand{\dataNumAnimals}{113 }
\newcommand{\dataNumPairsMeasured}{23,427 }
\newcommand{\dataMaxTimesMeasured}{59 }
\newcommand{\dataPercentPairsMeasured}{66}
\newcommand{\dataNumSigPairs}{1,314 }
\newcommand{\dataPercentSigPairs}{6 }

\newcommand{\dataFracInhibitory}{11 }

\newcommand{\dataPercentRecordMinOnce}{99}
\newcommand{\dataNumberRecordMinOnce}{186 }

\title{Neural signal propagation atlas of \textit{C. elegans}} %

\author[1]{Francesco Randi}
\author[1]{Anuj K Sharma}
\author[1]{Sophie Dvali}
\author[1,2,*]{Andrew M Leifer}
\affil[1]{Princeton University, Department of Physics, Princeton, NJ, 08544, United States of America}
\affil[2]{Princeton University, Princeton Neurosciences Institute, Princeton, NJ, 08544, United States of America}

\affil[*]{leifer@princeton.edu}

\begin{abstract}
A fundamental problem in neuroscience is understanding how a network's properties dictate its function. 
Connectomics provides one avenue to predict  nervous system function.  To test this explicitly, we systematically measure signal propagation in \dataNumPairsMeasured pairs of neurons across the head of the nematode \textit{Caenorhabditis elegans} by direct optogenetic activation and simultaneous whole-brain calcium imaging.
We measure the sign (excitatory or inhibitory), strength, temporal properties, and causal direction of signal propagation between these neurons to create a functional atlas. We find that signal propagation differs from predictions based on anatomy.
Using mutants, we show that  extrasynaptic signaling not visible from anatomy contributes to this difference. We identify many instances of dense-core-vesicle dependent signaling on seconds-or-less timescales that evoke acute calcium transients— often where no direct wired connection exists but where relevant neuropeptides and receptors are expressed.  We propose that here extrasynaptically released neuropeptides serve a similar function as that of classical neurotransmitters.
Finally, our measured signal propagation atlas better predicts neural dynamics of spontaneous activity than does anatomy. 
We conclude that both synaptic and extrasynaptic signaling drive neural dynamics on short timescales and that  measurement of evoked signal propagation  are critical for interpreting neural function. 
\end{abstract}
\begin{document}

\flushbottom
\maketitle

\thispagestyle{empty}

\section*{Main text}

Brain connectivity mapping is motivated by the claim that ``nothing defines the function of a neuron more faithfully than the nature of its inputs and outputs'' \cite{mesulam_imaging_2005,seung_towards_2011}. 
Connectomics measures inputs and outputs anatomically by mapping electrical and chemical synapses of the brain.  
It has been argued that the connectome is the primary description of a nervous system from which neural function and computation are derived. This view motivates large-scale efforts to measure  connectomes of a diverse set of organisms including mice \cite{abbott_mind_2020, helmstaedter_connectomic_2013}, fish \cite{hildebrand_whole-brain_2017}, flies \cite{scheffer_connectome_2020,dorkenwald_flywire_2022, schneider-mizell_quantitative_2016, eichler_complete_2017}, nematodes \cite{white_structure_1976, cook_whole-animal_2019,witvliet_connectomes_2021}, ascidians \cite{ryan_cns_2016} and platynereis \cite{veraszto_whole-animal_2020}. Partial  connectomes have already elucidated important details of direction selectivity in the mouse retina \cite{briggman_wiring_2011,bock_network_2011} and of the head direction system in \textit{Drosophila} \cite{hulse_connectome_2021,kim_ring_2017}. The \textit{C. elegans} connectome \cite{white_structure_1976,cook_whole-animal_2019,witvliet_connectomes_2021} is the most mature of all of these efforts, and it has helped reveal circuit-level mechanisms of   sensorimotor processing   \cite{chalfie_neural_1985, gray_circuit_2005,perkins_mutant_1986}, and is used to  constrain models of neural dynamics \cite{kunert-graf_multistability_2017,mi_connectome-constrained_2022} or to make predictions of neural function \cite{yan_network_2017}. 

An anatomical map of synaptic contacts, however, leaves ambiguous important aspects of neurons' inputs and outputs. A neural connection's strength and sign (excitatory or inhibitory) is not always evident from anatomy or gene expression. 
Many mammalian neurons release  both excitatory and inhibitory neurotransmitters, thus requiring functional measurements to disambiguate \cite{vaaga_dual-transmitter_2014}.   For example,  starburst amacrine cells release both GABA and acetylcholine  \cite{omalley_co-release_1992}; serotonergic neurons in the dorsal raphe nucleus also release glutamate \cite{johnson_synaptic_1994}; and neurons in the ventral tegmental area  release two or more of dopamine, GABA and glutamate \cite{yoo_ventral_2016}. 
Furthermore, anatomy does not provide a clear picture of the timescales of neural transmission, and not all anatomical neural connections are operational. In the head compass circuit in \textit{Drosophila}, many anatomical connections exist, but plasticity from long-term potentiation and long-term depression selects only a subset to function in order to guide relevant visual inputs\cite{fisher_sensorimotor_2019}. Similarly functional connections in the central complex  appear to be sparser than expected from anatomy \cite{franconville_building_2018}. Neuromodulators also  adjust  properties of neural connections that aren't visible from anatomy in order  to  strengthen or weaken them or to turn on only a subset of circuits out of a larger menu of possible latent circuits, for example,  in the crab stomatogastric ganglion \cite{harris-warrick_modulation_1991,marder_neuromodulation_2012,bargmann_beyond_2012}. Additionally, neurons  can release signaling  molecules from outside the synapses,  to facilitate extrasynaptic neural connections that are not visible in the anatomical wiring \cite{bentley_multilayer_2016}, as we explore in this work. These additional properties of a neuron's inputs and outputs  pose challenges for accurately predicting network  function from anatomy alone. 

A more direct way to characterize neural inputs and outputs is to measure their functional properties directly by activating a neuron and observing responses in other neurons, what we refer to as  neural signal propagation. 
Indeed, complementing anatomical measurements with functional measurements  was critical for understanding   direction selective circuits in the mouse retina \cite{briggman_wiring_2011, bock_network_2011}.  
Measuring signal propagation  captures the strength and sign of neural connections, and it reflects  plasticity, neuromodulation, and even extrasynaptic signaling. Moreover, direct measures of signal propagation allow us to define mathematical relations that describe how neural activity of an upstream neuron drives activity in a downstream neuron including its temporal response profile.
Historically these measurements have been variously called,
``influence mapping'' \cite{panzeri_structures_2022}, ``projective fields''  \cite{lehky_network_1988}, ``evoked brain connectivity'' \cite{lepage_inferring_2013} and ``functional connectivity'' \cite{franconville_building_2018}.    
This  contrasts with  approaches that rely on correlations observed in neural activity to infer communications between neurons (confusingly also referred to as ``functional connectivity''). Correlative approaches do not directly observe causality, nor do they directly capture detailed temporal properties of signal transmission, and they are limited to finding relations among only those neurons that happen to be active. By instead measuring downstream responses  to neural activation, we  probe connections between even those neurons that are not spontaneously active.

Previous efforts have measured signal propagation using   optogenetic perturbations \cite{petreanu_channelrhodopsin-2assisted_2007, huber_sparse_2008} and calcium imaging \cite{zhang_optical_2007, emiliani_all-optical_2015}  \textit{in vivo}   in mouse visual cortex \cite{wilson_division_2012}, mouse hippocampus \cite{rickgauer_simultaneous_2014}, mouse somatosensory cortex \cite{packer_simultaneous_2015},  zebrafish \cite{mcraven_high-throughput_2020},  \textit{Drosophila} \cite{chen_functional_2021}, and \textit{C. elegans} \cite{guo_optical_2009}. A  cell-type resolution functional  map of  the  \textit{Drosophila} central complex brain region  was  used to relate functional  measurements via neural activation to anatomy \cite{franconville_building_2018}. All of these prior works  were restricted to selected circuits or subregions of the brain, and many achieved only cell-type and not single-cell resolution.

Here we use neural activation to make functional measurements of  signal propagation between neurons throughout the  head of \textit{C. elegans}  at single-cell resolution. We survey \dataNumPairsMeasured neuron pairs to present a systematic signal propagation atlas. We show that functional measurements better predict  spontaneous activity than  anatomy and gene expression, and that peptidergic extrasynaptic signaling  contributes to neural dynamics by performing a functional role similar to that of a classical neurotransmitter. 

\section*{Results}
\subsection*{Simultaneous population imaging and single-cell activation}
\begin{figure}[htbp]
\centering
\includegraphics[width=.9\linewidth]{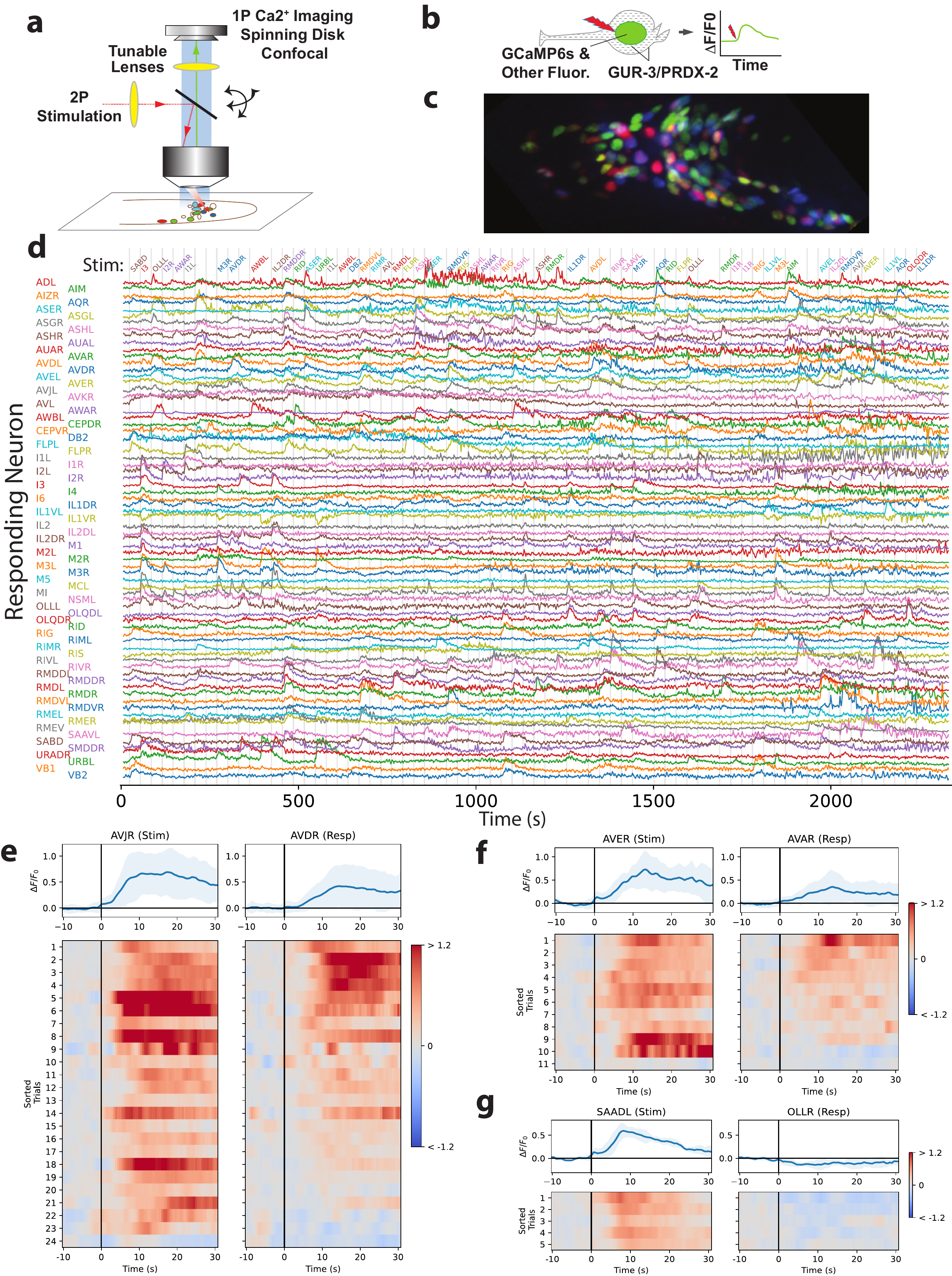}
\caption{\textbf{Measuring neural activation and network response.} Schematics of  \textbf{a)} the instrument  and \textbf{b)} of the experiment. \textbf{c)} NeuroPAL fluorophores for neural identification. \textbf{d)} Whole-brain cell-resolved calcium activity (GCaMP6s fluorescence normalized by noise) during stimulation of individual neurons (gray vertical lines, listed at the top). \textbf{e)} Paired activity of  AVJR and AVDR in response to AVJR stimulation, shown as relative change $\Delta F/F_0$. Top panel: average (blue) and standard deviation (shading) across trials and animals. Bottom panel: simultaneously recorded paired activity for individual trials (sorted by mean AVDR activity). All trials are shown that elicitced activity. \textbf{f)} Same as \textbf{e} for  AVER stimulation and AVAR response. \textbf{ g)}  Same as \textbf{e} for SAADL stimulation and OLLR  response. }  \label{fig:demonstration}
\end{figure}

To   measure signal propagation, we  activated each single neuron, one at a time, while simultaneously recording population calcium activity at cellular resolution. We recorded  activity from \dataNumAnimals wild-type (WT) background animals, each for up to 40 min, while stimulating a mostly-randomly selected sequence of neurons one-by-one every 30 s (Fig.~\ref{fig:demonstration}).
We  combined whole-brain  calcium imaging via spinning disk single-photon confocal microscopy \cite{nguyen_whole-brain_2016, venkatachalam_pan-neuronal_2016} with two-photon \cite{denk_two-photon_1990} targeted optogenetic stimulation \cite{rickgauer_two-photon_2009}, each with their own remote focusing system, to measure and manipulate neural activity in an immobilized animal (Fig.~\ref{fig:demonstration}a). Animals were awake and pharyngeal pumping was visible during recordings. We spatially restricted the optogenetic excitation volume in three dimensions to the typical size of a \textit{C. elegans} cell soma (Extended Data Fig.~\sfigspotwavelength a) using temporal focusing  \cite{andrasfalvy_two-photon_2010, rickgauer_simultaneous_2014} to address single neurons without also activating their neighbors (Extended Data  Figs.~\sfigbleaching{}c,d).  

To overcome challenges associated with spectral overlap \cite{rickgauer_simultaneous_2014, packer_simultaneous_2015, yang_simultaneous_2018, zhang_closed-loop_2018, russell_all-optical_2022}, we expressed the GUR-3/PRDX-2 purple-light activatable optogenetic system \cite{bhatla_light_2015, quintin_distinct_2022}  and a nuclear-localized  calcium indicator GCaMP6s \cite{chen_ultrasensitive_2013} in each neuron (Fig.~\ref{fig:demonstration}b). 
Light activation of GUR-3 had previously been shown to elicit both calcium responses and behavioral responses. For example, light-activation of GUR-3 in I2 was shown to inhibit pharyngeal pumping by release of glutamate \cite{bhatla_distinct_2015}. 
Light activation of GUR-3/PRDX-2 expressed in AVA evoked reversals (Extended Data Fig.~\ref{ext:stim_characterization}h),  as expected \cite{gordus_feedback_2015, li_encoding_2014}.
To obtain high expression levels of GUR-3/PRDX-2 with less toxicity we expressed  it under the control of a drug-inducible gene expression system and only turned on gene expression prior to experiments.
The stimulus illumination duration of 0.3 s or 0.5 s was chosen to evoke modest  amplitude calcium responses (Extended Data Fig.~\ref{sfig:autoresponse_amplitude}f), similar in amplitude to those evoked by natural odor stimuli \cite{lin_functional_2023}.
Activation of cholinergic motor neurons M1 under these stimulation conditions evoked  pharyngeal muscle contraction (Supplementary Video 1).
This provides evidence that our stimulation condition   is well suited to evoke typical synaptic release of classical neurotransmitters,
since M1 is known to release acetylcholine via chemical synapse onto  pharyngeal muscles \cite{franks_comparison_2009, sando_hourglass_2021}.

We performed calcium imaging, with excitation light at a wavelength and intensity  that does not elicit  photoactivation of GUR-3/PRDX-2 (Extended Data Fig.~\sfigspotwavelength b) \cite{bhatla_c._2009}. We also used genetically encoded fluorophores from  NeuroPAL expressed in each neuron \cite{yemini_neuropal_2021}  to identify neurons  consistently across animals (Fig.~\ref{fig:demonstration}c).  %
 Many neurons exhibited calcium activity in response to activation of one or more other neurons (Fig.~\ref{fig:demonstration}d). A downstream neuron's response to a stimulated neuron is evidence %
that a signal propagated from the stimulated neuron to the downstream neuron.

We highlight three examples from the motor circuit (Fig.~\ref{fig:demonstration}e-g).   
Stimulation of the interneuron AVJR evoked activity in AVDR (Fig.~\ref{fig:demonstration}e).   AVJ  had been predicted to  coordinate   locomotion  upon egg laying by promoting forward movements  \cite{hardaker_serotonin_2001}.   AVD activity is associated with sensory-evoked (but not spontaneous) backward locomotion  \cite{chalfie_neural_1985, wicks_dynamic_1996, gray_circuit_2005,  kawano_imbalancing_2011} %
and  receives chemical and electrical synaptic input from AVJ  \cite{white_structure_1986,witvliet_connectomes_2021}. %
Both the wiring and our functional measurements suggest that AVJ may also play a role in coordinating backward locomotion, in addition to its previously described roles related to egg laying and forward locomotion.

Activation of premotor interneuron  AVE evoked activity transients in AVA (Fig.~\ref{fig:demonstration}f).  %
Both  AVA \cite{chronis_microfluidics_2007, guo_optical_2009, arous_automated_2010, kawano_imbalancing_2011, shipley_simultaneous_2014, kato_global_2015, wang_flexible_2020, faumont_image-free_2011} (Extended Data Fig.~\ref{ext:stim_characterization}h)  and AVE \cite{kawano_imbalancing_2011, wang_flexible_2020}  are implicated in backward movement and have activity correlated with one another \cite{kawano_imbalancing_2011} and AVE makes gap junction and  many chemical synaptic contacts with AVA \cite{white_structure_1986, witvliet_connectomes_2021}.  

Activation of the turning-associated neuron SAADL \cite{wang_flexible_2020} inhibited activity in sensory neuron OLLR.  SAAD had been predicted to inhibit OLL based on gene expression measurements \cite{fenyves_synaptic_2020}. SAAD is cholinergic and it makes chemical synapses to OLL which expresses an acetylcholine-gated chloride channel, LGC-47 \cite{taylor_molecular_2021, jones_cys-loop_2008, witvliet_connectomes_2021}.

Selected additional neural responses that are consistent with previous reports in the literature are listed in Extended Data Fig.~\ref{ed:agreement_table}.

\subsection*{\Themap}
\begin{figure}[htbp]
\centering
\includegraphics[page=1, width=.85\linewidth]{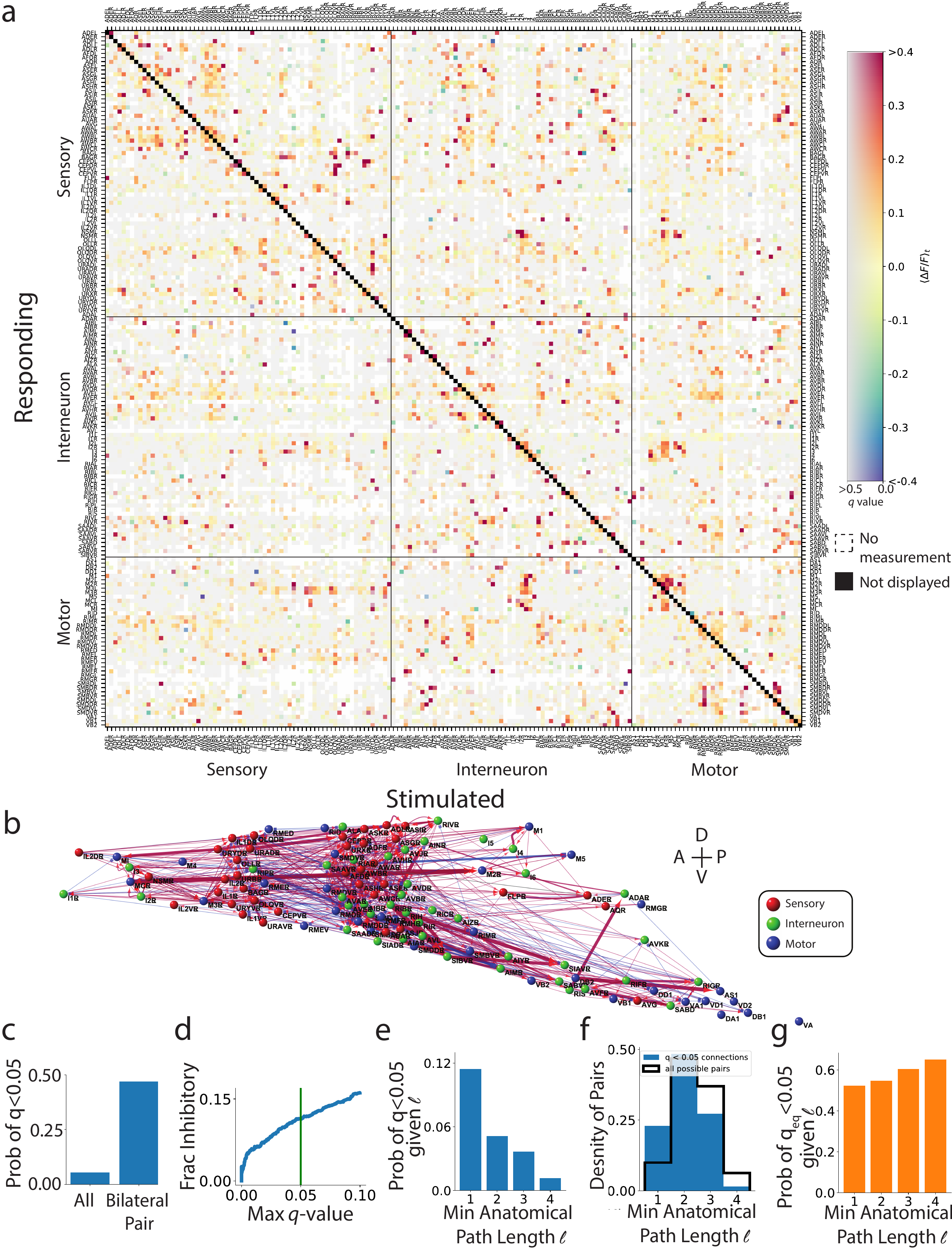}
\caption{ \textbf {\Themap{} of \textit{C. elegans}. a)} Mean post-stimulus neural activity $\langle \Delta F/F_0 \rangle_t$ averaged across trials and individuals. $q$-values report false discovery rate (more gray is less significant). White indicates no measurement. Auto-response is required for inclusion and not displayed (black diagonal). $N$=\dataNumAnimals animals. Neurons that were recorded but never stimulated are shown in Extended Data Fig.~\ref{sfig:maps_complete}. \textbf{b)} Corresponding network graph with neurons positioned anatomically (only $q<0.05$ connections). Width and transparency indicate mean response amplitude (red, excitatory; blue, inhibitory). \textbf{c)}  A bilaterally symmetric pair is more likely to have a $q<0.05$ functional connection than a pair chosen at random. \textbf{d)}  Fraction of  connections that are inhibitory as a function of the $q$-value threshold. Green indicates $q<0.05$.  \textbf{e)}  Probability of being functionally connected ($q<0.05$) given minimum anatomical path length $l$. \textbf{f)}  Distribution of $l$ for functionally connected pairs (blue) compared to all possible pairs (black). \textbf{g)} Probability of being functionally non-connected  ($q_\mathrm{eq}<0.05$) given $l$. }  \label{fig:map}
\end{figure}

We generated a \themap{}  by aggregating downstream responses to stimulation for each neuron pair across  recordings from \dataNumAnimals individuals (Fig.~\ref{fig:map}a). We report the   average calcium response in a time window $\langle\Delta F/F_0\rangle_t$ averaged across  trials and animals (Extended Data Fig.~\sfigDFonly a). 
We imaged  activity in response to stimulation for \dataNumPairsMeasured neuron pairs  (\dataPercentPairsMeasured\% of all possible pairs in the head)  at least once, and  as many as \dataMaxTimesMeasured times (Extended Data Fig.~\ref{sfig:num_obs_matrix}a). This includes activity from \dataNumberRecordMinOnce of 188 neurons in the head (\dataPercentRecordMinOnce\% of all head neurons, Extended Data Fig.~\ref{sfig:num_obs_hist}b).

We employed two statistical tests to identify neuron pairs that  are ``functionally connected,''  `functionally non-connected'' or for which we lack confidence to make either determination. Both tests compare observed  calcium transients  in each downstream neuron to a null distribution of transients  recorded in  experiments lacking stimulation. To test whether a neuron pair is functionally connected, we seek to reject the null hypothesis that its downstream calcium transients could have arisen from the null distribution. We report a $q$-value, related to significance, that conveys the false-discovery rate across the many hypotheses tested in our large dataset\cite{benjamini_controlling_1995, storey_statistical_2003} (Extended Data Fig.~\ref{sfig:q_only}b).  Pairs that achieve $q<0.05$ are deemed ``functionally connected.'' To test whether a neuron pair is functionally non-connected, we perform a separate, equivalence test that seeks to reject the null hypothesis that the effect size of the observed transient is larger than some small $\epsilon$. Here too we report a false discovery rate, $q_{\mathrm{eq}}$ (Extended Data Fig.~\ref{sfig:q_eq}b). Neuron pairs with  $q_{\mathrm{eq}}<0.05$ are declared functionally non-connected.

We emphasize that this statistical framework is conservative and presents a very high bar for which to declare a neuron pair to be functionally connected  or functionally non-connected. Passing either of these hurdles requires consistent and reliable responses (or non-responses) and takes into account effect size, sample size, and multiple hypothesis testing.  The majority of neuron pairs we measure fail to pass either of these tests, even though they often contain neural activity that, when observed in isolation, could  easily be classified as a response  (e.g.~AVJR->ASGR in Extended Data Fig.~\ref{sfig:q}c).

Our \themap ~comprises the  response amplitude and its associated $q$-value  (Fig.~\ref{fig:map}a, Extended Data Fig.~\ref{sfig:maps_complete}a).  
  This functional dataset,  overlaid on the anatomical wiring diagram \cite{white_structure_1986, witvliet_connectomes_2021}, is browseable  online (\url{https://funconn.princeton.edu}) via software  built on the NemaNode platform \cite{witvliet_connectomes_2021}.
We estimate that at  least \dataNumSigPairs of the \dataNumPairsMeasured measured neuron pairs, or \dataPercentSigPairs{}\%, pass our stringent criteria to be deemed functionally connected at $q<0.05$  (Fig.~\ref{fig:map}c). We also mapped our confidence in neuron pairs that are functionally non-connected (Extended Data Fig.~\ref{sfig:q_eq}b).
Note that in all cases  functional connections refer to   ``effective connections''  because they represent the propagation of signals over all paths in the network between the stimulated and responding neuron, not just the direct (monosynaptic) connections between them. 

\textit{C. elegans} neuron subtypes typically consist of two bilaterally symmetric neurons, often connected by gap junctions,  that  have similar neural wiring \cite{white_structure_1986}, similar gene expression \cite{taylor_molecular_2021}, and correlated activity \cite{uzel_set_2022}. Our measurements show that bilaterally symmetric neurons are eight times more likely to be functionally connected than pairs of neurons chosen at random (Fig.~\ref{fig:map}c).  

The balance of excitation and inhibition  is important for a  network's stability \cite{van_vreeswijk_chaos_1996, isaacson_how_2011} but until now has not been directly measured in the worm. 
We measure that \dataFracInhibitory{}\% of  $q<0.05$ functional connections are inhibitory (Fig.~\ref{fig:map}d), comparable to prior estimates of  $\approx20\%$ of synaptic contacts in \textit{C. elegans} \cite{fenyves_synaptic_2020} or  $\approx20\%$ of cells in mammalian cortex \cite{meinecke_gaba_1987}. Our estimate is likely a lower bound, because we assume that we only observe inhibition in neurons that already have tonic activity.

Neurons with a single-hop anatomical connection were more  likely to be functionally connected at $q<0.05$ compared to neurons with only indirect or multi-hop anatomical connections; and functional connections became less likely as the minimal path length through the anatomical network increased  (Fig.~\ref{fig:map}e).
Conversely, neurons that had large minimal  path-lengths through the anatomical network were more likely to be functionally non-connected than neurons that had a single-hop minimal path length  (Fig.~\ref{fig:map}g). 
We investigated how far   responses to neural stimulation penetrate into the anatomical network. 
Functionally connected ($q<0.05$) neurons   were on average  connected by a minimal anatomical path length of 2.1 hops (Fig.~\ref{fig:map}f),
suggesting   that neural perturbations often propagate multiple hops through the anatomical network or that neurons are also signaling directly through non-wired means, as explored later. 

We  observed instances of two types of variability in neural responses  across  trials and individuals: (1) a downstream neuron  responds to stimulation of a given upstream neuron in only some simulations but not  others (Extended Data Fig.~\ref{sfig:variability}a) and (2) the amplitude, temporal shape, and sign of the response varies from one stimulation to the next (Extended Data Fig.~\ref{sfig:variability}b-e). Some variability in the responding neuron's activity  can be attributed to variability of the upstream neuron's activity.
(We refer to a neuron's response to its own stimulation as an auto-response.) To study variability in the strength, temporal properties, and sign of the connection, while excluding variability contributed by the upstream neuron's activity, we calculated a  kernel for each stimulation that caused a response. The kernel gives the activity of the downstream neuron when convolved with the activity of the upstream neuron.
The kernel describes how the signal is transformed  from   upstream to   downstream neuron for that stimulus event, including the timescales of the signal transfer (Extended Data Fig.~\sfigkernels b,c). We characterized  variability of each functional connection  by comparing how these kernels transform a standard stimulus  (Extended Data Fig.~\ref{sfig:variability}e).  Many neuron pairs had collections of kernels with properties that varied across trials. We did not  identify the sources of  this variability, but they likely include state- and history-dependent effects \cite{gordus_feedback_2015}, including from neuromodulation \cite{harris-warrick_modulation_1991,stern_neuromodulatory_2017}, plasticity, and inter-animal variability in wiring and expression.  Collections of kernels within a neuron  pair were more stereotyped than  collections of kernels randomly selected from across shuffled  pairs (Extended Data Fig.~\ref{sfig:variability}f), as expected.

\subsection*{Functional connectivity differs from anatomy}
\begin{figure}[htbp]
\centering
\includegraphics[page=1, width=1\linewidth]{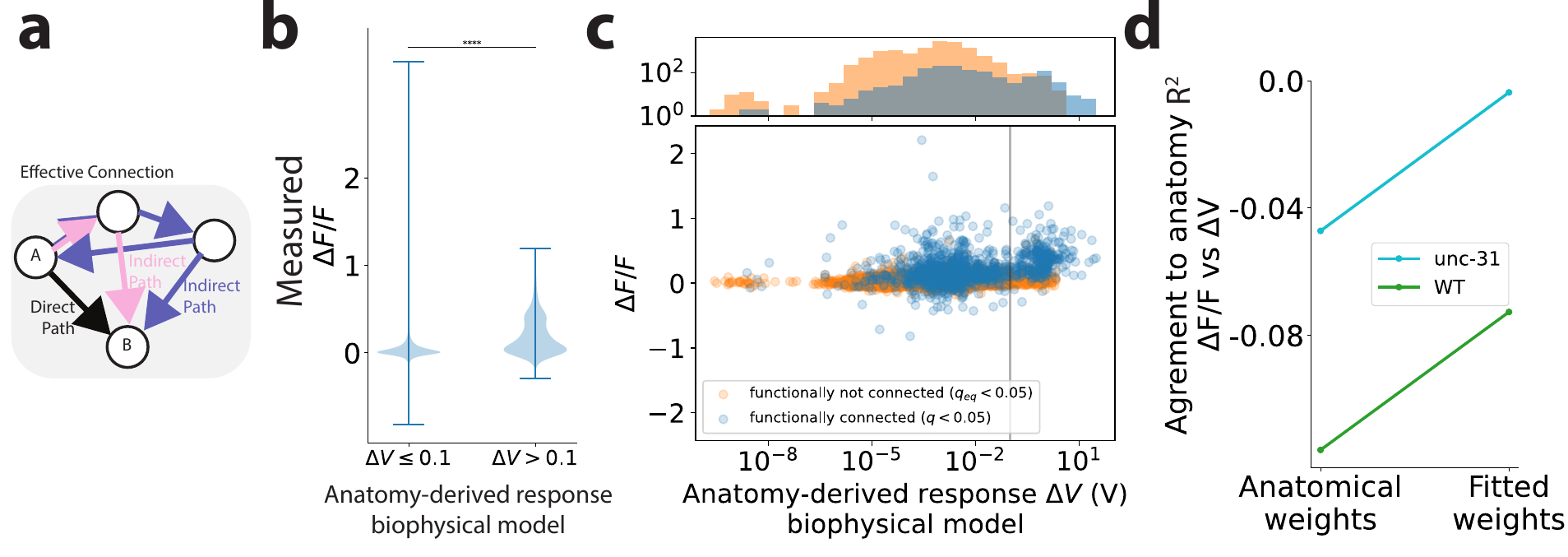}
\caption{\textbf{Functional connectivity differs from anatomy.} \textbf{a)} Functional connectivity describes effective connections, which can include direct (black), indirect (pink), and recursive (purple) paths. Instead, anatomical features like synapse count describe only direct paths (black). To make a like-to-like comparison, anatomy-derived effective connections are computed in connectome-constrained simulations. \textbf{b)} Pairs predicted from anatomy to have large stimulus evoked downstream responses ($\Delta V>0.1V$) tend to have stronger measured responses (larger $\Delta F/F$ in the downstream neuron) than pairs that anatomy predicts to have small stimulus-evoked downstream responses. ($p<0.0001$, one-sided KS test). \textbf{c)} Main panel: Measured downstream response ($\Delta F/F$) vs. anatomy-derived response ($\Delta V$) for pairs that we observe to be functionally connected ($q<0.05$, blue) and functionally non-connected ($q_\mathrm{eq}<0.05$, orange). The vertical gray line is the threshold ($0.1 V$) that separates anatomy-predicted effective connections from non-connections. Top panel: Marginal distributions (note y-axis is log scale). Functionally connected pairs are enriched for anatomy-predicted effective connections, compared to functionally non-connected pairs. ($p<0.0001$, one-sided KS test). \textbf{d)} Agreement of measured responses to anatomy-predicted responses is shown for wild-type (green) and \textit{unc-31} animals (cyan) in two conditions: using weights and signs  from anatomy, or when weights and signs are fitted optimally. Agreement is calculated as the $R^2$ coefficient for the line-of-best-fit: $\Delta F/F = m \Delta V$. Perfect agreement would be $R^2=1$.
} 
\label{fig:anatomy}
\end{figure}

We sought to compare our measured functional connectivity to  anatomy  \cite{white_structure_1986,varshney_structural_2011,witvliet_connectomes_2021}.   Functional connectivity and anatomical connectivity describe different levels of the network -- functional connectivity  measures   effective connection  between two neurons, including   contributions from all paths through the network, direct and indirect  (Fig.~\ref{fig:anatomy}a). In contrast, anatomical features such as synapse count are properties of only the direct (monosynpatic) connection between  two neurons. A bridge is needed between these two levels of description in order to make a like-to-like comparison. Therefore, we used  connectome-constrained biophysical models of the network to simulate the expected signal propagation based on anatomy. We  activated neurons \textit{in silico} and  simulated  the network's response, using synaptic weights from anatomy \cite{white_structure_1986, witvliet_connectomes_2021},   synaptic polarities estimated from gene expression   \cite{fenyves_synaptic_2020},  and  using common assumptions about timescales and dynamics \cite{kunert_low-dimensional_2014}.  

The anatomy-derived biophysical model made some predictions that agreed with our functional measurements.
We classified neuron pairs into  anatomy-predicted effective connections or anatomy-predicted effective non-connections based on the  biophysical model (threshold of $\Delta V=0.1$). 
Anatomy-predicted effective connections were significantly more likely to correspond to larger amplitude responses in our measurements than predicted non-connections (Fig.~\ref{fig:anatomy}b), showing agreement between structure and function. 
Similarly, we find that functionally connected pairs of neurons ($q<0.05$) are relatively enriched for anatomy-predicted effective connections compared to functionally non-connected neurons  ($q_{\mathrm{eq}}<0.05$),  (Fig.~\ref{fig:anatomy}c, top panel). 

Overall, however, there was fairly poor agreement between anatomical predictions and measurement. For example, we measured non-zero responses in neuron pairs that were predicted from anatomy to have almost no response (Fig.~\ref{fig:anatomy}c).  The agreement between anatomy predicted and measured response was also poor when considering all neuron pairs  (Fig.~\ref{fig:anatomy}d, $R^2$<0, where an $R^2$ of 1 would indicate perfect agreement).

It is challenging to infer the strength, sign or other functional properties of a neural connection from anatomy and this may contribute to the disagreement between anatomical predictions and our measurements. In  mammals \cite{vaaga_dual-transmitter_2014}  and worms \cite{fenyves_synaptic_2020} presynaptic neurons can send both excitatory and  inhibitory signals to their postsynaptic partner  leaving the overall strength and sign ambiguous. For example, the  AFD-AIY pair expresses  machinery for inhibition via glutamate but  we and others find it to  be functionally excitatory likely due to peptidergic signaling \cite{narayan_transfer_2011} (Extended Data Fig.~\ref{sfig:AFD-AIY}g). ASE-AIB is another ambiguous pair reported in the literature \cite{kuramochi_excitatoryinhibitory_2019}. 

We therefore wondered whether agreement between structure and function would improve if we allowed the strength and signs of our biophysical model to float, but forbade the creation of entirely new connections that hadn't appeared in the connectome. We fit strength and sign in the biophysical model  from information about the measured effective functional connections. For simplicity during fitting, we assumed a linear network at steady state, but during  the  comparison we relaxed this assumption.  
Allowing the anatomical weights and signs to change in the most favorable way --but without adding any new connections --  improved agreement between the anatomical prediction and functional measurements, although overall agreement remained poor  (Fig.~\ref{fig:anatomy}d). 
 
 We therefore explored whether additional functional connections  exist that are not present in the anatomical wiring. We measured signal propagation in \textit{unc-31} mutant animal defective for dense-core vesicle-mediated extrasynaptic signaling, as explained below. While agreement was still poor,  the signal propagation in these animals showed better agreement with anatomy  than for WT (Fig.~\ref{fig:anatomy}d). This prompted us to explore extrasynaptic signaling further.

\subsection*{Extrasynaptic signaling also drives neural dynamics}
Neurons can communicate extrasynaptically by releasing transmitter that binds to receptors of downstream neurons after diffusing through the extracellular milieu instead of directly traversing a synaptic cleft.  Such extrasynaptic signaling, sometimes called ``volume transmission'' \cite{agnati_intercellular_1995},   occurs either because the transmitter is released far from the cleft, for example via dense core vesicle, or because transmitter released at the cleft spills out into the milieu. Extrasynaptic signaling  has been reported for  GABA \cite{brickley_extrasynaptic_2012, shen_extrasynaptic_2016}, NMDA \cite{hardingham_synaptic_2010}, monoamines \cite{bentley_multilayer_2016}, and neuropeptides \cite{taghert_peptide_2012}.  In all cases, extrasynaptic signaling is not visible from anatomy  and therefore forms additional wireless layers of communication in the  nervous system   \cite{bentley_multilayer_2016}. We are  motivated to investigate extrasyaptic signaling of neuropeptides in particular because neuropeptides and neuropeptide receptors are  ubiquitous across not only the \textit{C. elegans} nervous system \cite{taylor_molecular_2021,ripoll-sanchez_neuropeptidergic_2022,beets_system-wide_2022} but also in mammalian cortex \cite{smith_single-cell_2019}. Neuropeptides are typically released via dense core vesicles \cite{pelletier_electron_1977} and are not required to be released at the synaptic cleft. Here we refer to dense-core-vesicle mediated signaling as extrasynaptic because it is commonly observed far from the synapse, however we cannot exclude the possibility that dense-core vesicles may also be released at the synapse.

To probe the role of dense-core mediated extrasynaptic signaling we measured a \themap{}
of \textit{unc-31}-mutant animals defective for dense-core vesicle-mediated release (Extended Data Fig.~\ref{sfig:maps_unc31}a, 18 individuals)
and compared their neural responses to WT animals. 
This mutant disrupts dense-core vesicle-mediated extrasynaptic signaling of peptides and monoamines because it lacks the UNC-31/CAPS protein that is  involved in  dense core vesicle fusion.  In  \textit{C. elegans}, defects in UNC-31/CAPS are not expected to disrupt chemical or electrical synapses \cite{speese_unc-31_2007}. The \themap{} of this mutant is browseable online and can be compared   to WT at \url{https://funconn.princeton.edu}.

Extrasynaptic signaling is  often associated with neuromodulation and is assumed to alter excitability or modulate synaptic properties that change how neurons respond to inputs  \cite{harris-warrick_modulation_1991, taghert_peptide_2012} often over long timescales, though not always \cite{golowasch_proctolin_1992}.  But extrasynpatic transmission can also directly evoke activity in downstream neurons. Our signal propagation measurements are not designed to directly detect changes to a neuron's excitability and  instead such changes might appear as variability in our measured responses. But our measurements should detect instances where extrasynaptic signaling evokes activity. 

Many functional connections remain in the \textit{unc-31} mutant, consistent with our expectation that chemical synapse and gap-junction mediated signaling plays a prominent role in the nervous system (Extended Data Fig.~\sfiguncvsWTcontrol). 
For neuron pairs observed in both WT and \textit{unc-31} mutants, \textit{unc-31} animals had a smaller proportion of  high-confidence functional connections (Extended Data Fig.~\sfiguncvsWTstats b), consistent with the expected loss of  extrasynaptic \textit{unc-31}-dependent signaling between neurons.  %

\begin{figure}[htbp]
\centering
\includegraphics[page=1, width=1\linewidth]{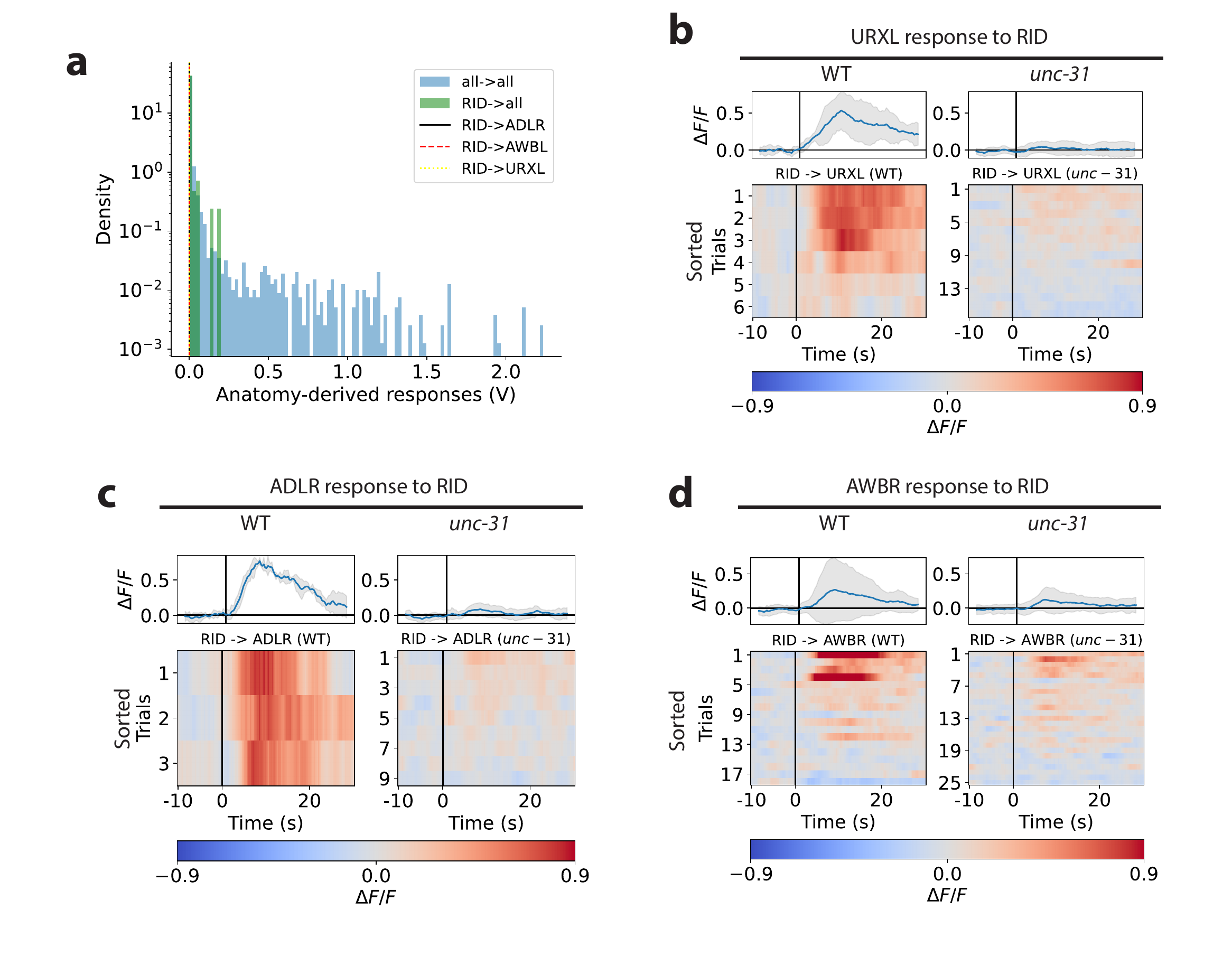}
\caption{\textbf{Anatomy omits extrasynaptic signaling from neuron RID. a)} ADL, AWB, and URX are among the neurons predicted from anatomy  to have no response to RID stimulation  because there is no strong anatomical path from RID to those neurons (vertical lines at or near 0 volts). Their RID-evoked anatomy-predicted responses are shown within the distribution  of anatomy-predicted responses for all neuron pairs (blue histogram), as in Fig.~\ref{fig:anatomy}b. 
\textbf{b-d.)} Activity of  neurons \textbf{b)} URXL, \textbf{c)} ADLR,  and \textbf{d)} AWBR to RID stimulation, in WT and \textit{unc-31} mutant backgrounds. Top panel: average across trials and animals. Bottom panel: individual traces are sorted across trial and animal by mean response amplitude. Here trials are shown even in cases when RID activity was not measured. Responses of additional neurons are shown in Extended Data Fig.~\ref{sfig:RID_more_responses}c }  \label{fig:wireless}
\end{figure}

We sought to investigate  specific extrasynaptic connections because these connections contribute to  discrepancies between an anatomical and functional description. We  first turned to the neuron RID, a neuroendocrine-like cell  that is thought to  signal to other neurons extrasynaptically via neuropeptides such as FLP-14, PDF-1, and  INS-17   \cite{lim_neuroendocrine_2016, taylor_molecular_2021}. 
RID has many potential extrasynaptic signaling partners  but only very few and weak outgoing wired connections making it a good candidate in which to observe extrasynaptic signaling. In our imaging strain, RID exhibited only dim tagRFP-T expression, which prevented us from consistently segmenting the neuron in order to capture  RID's own calcium response to activation. We nonetheless stimulated RID and observed other neurons' responses (Extended Data Fig.~\sfigRIDmoreresponses c).  For analysis of  responses to RID activation in Extended Data Fig.~\sfigRIDmoreresponses c and Fig.~\ref{fig:wireless}a only, we have  relaxed our inclusion requirements and include downstream neural responses even when we do not measure the calcium activity of RID.

We inspected the activity of three neuron subtypes, URX, ADL, and AWB, that were predicted  to have little or no response to RID stimulation based on anatomy (Fig.~\ref{fig:wireless}a) but showed notably strong responses to RID stimulation when measured in WT background  (Fig.~\ref{fig:wireless}b-d). Several lines of evidence led us to conclude that RID predominantly sends signals to URX, ADL and AWB extrasynaptically.
(1) When RID was stimulated in the \textit{unc-31} background, these three neurons all exhibited reduced amplitude or were less likely to respond at all (Fig.~\ref{fig:wireless}b-d), suggesting that these neurons' connections to RID are dense-core-vesicle-dependent.
(2) All three neuronal subtypes  express receptors for peptides produced by RID (NPR-4 and NPR-11 for FLP-14 and PDFR-1 for PDF-1). And (3) there are no direct (monosynaptic) connections through the anatomical network from RID to URX, ADL, or AWB. 
The shortest paths from RID to URXL or AWBR require two hops ($l=2$), and three hops for ADLR ($l=3$), 
and in each case they rely on fragile single-contact  synapses   that  appear in only one out of the four individual connectomes \cite{witvliet_connectomes_2021}. 
Taken together, we conclude that RID signals to other neurons extrasynaptically and that this signalling is captured by our   functional measurements   but not by anatomy.  

\begin{figure}
    \centering
    \includegraphics{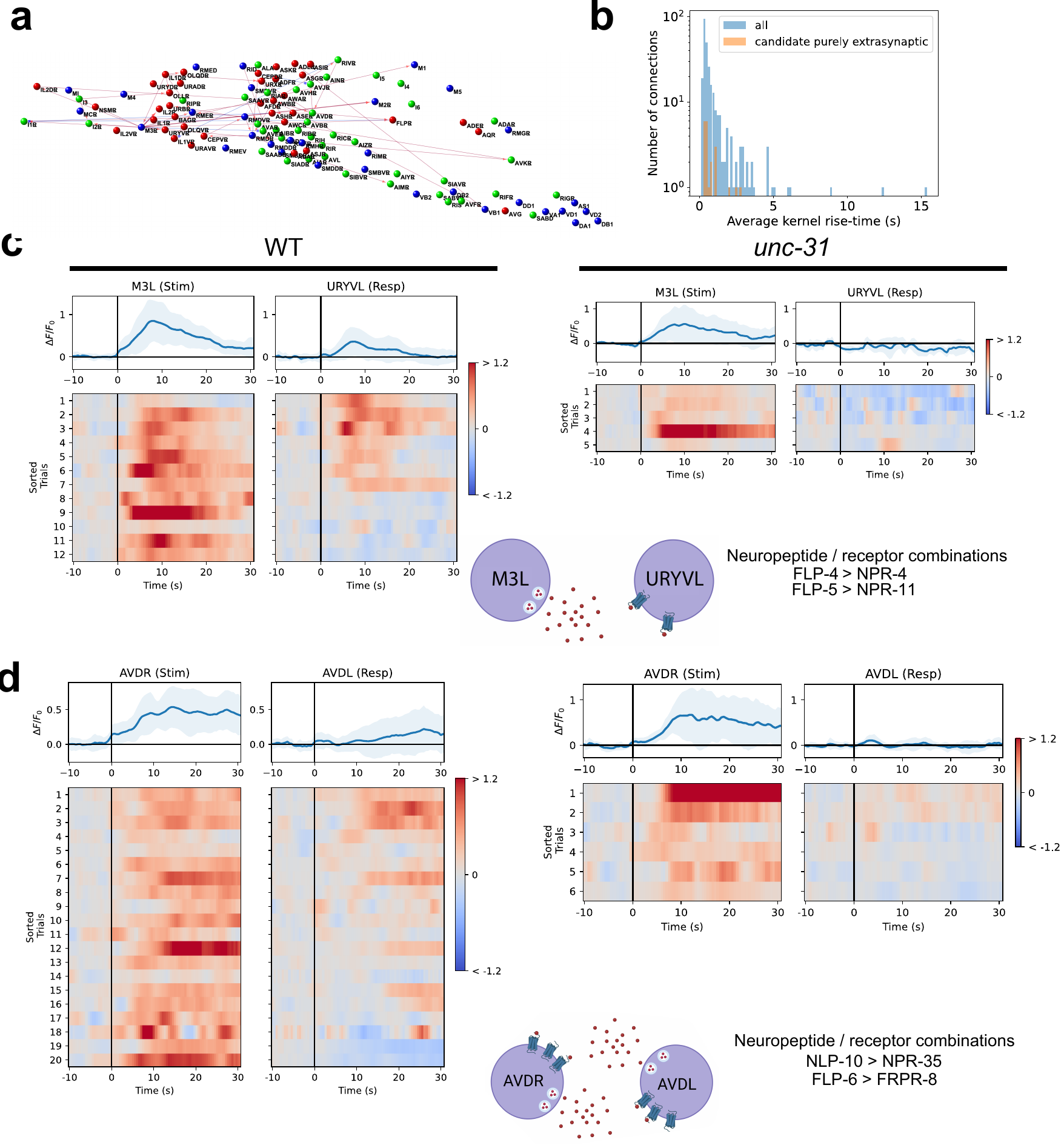}
    \caption{\textbf{a)} Candidate purely extrasynaptic-dependent connections.
        \textbf{b)} Distribution of kernel rise times, as in Extended Data Fig.~\ref{sfig:kernels}d.  \textbf{c)} Paired responses for  M3L->URYVL and \textbf{d)} AVDR->AVDL, respectively, shown for WT and \textit{unc-31} animals. \textit{unc-31} animals do not show downstream responses to stimulation.  AVDR->AVDL extrasynaptic communication is putatively mediated in autocrine loops via NLP-10->NPR-35 and FLP-6->FRPR-8 signaling.
        }
    \label{fig:esyn}
\end{figure}

\subsection*{A screen for purely extrasynaptic-dependent signal propagation}
Neuron RID had already been identified in the literature as a neuroendocrine-like cell that likely communicates extrasynaptically~\cite{lim_neuroendocrine_2016}. To  identify new pairs of neurons that are purely  extrasynaptic-dependent, we performed an unbiased screen  and selected for  neuron pairs  that had functional connections in wild-type animals ($q<0.05$) but were  functionally non-connected  in animals defective for extrasynaptic signaling ($q_{\mathrm{eq}}<0.05)$. Out of the \dataNumSigPairs pairs that we are confident are functionally connected in WT animals ($q<0.05$), 53 met the stringent threshold of also being functionally non-connected in  \textit{unc-31} animals (Fig.~\ref{fig:esyn}a, Extended Data Fig.~\ref{sfig:esyn}.  These putative purely extrasynaptic-dependent  connections represent a conservative lower-bound on the number of purely extrasynaptic-dependent connections in the brain. 

Notably the distribution of timescales of \textit{unc-31}-mediated functional connections is similar to that of all functional connections, Fig.~\ref{fig:esyn}b.

Neuron pair M3L->URYVL is a representative example of a  purely extrasynaptic-dependent connection found from our screen. There are no direct wired connections between M3L and URYVL, but stimulation of M3L evokes \textit{unc-31}-dependent calcium activity in URYVL (Fig.~\ref{fig:esyn}c).   The recent  de-orphaniziation of many neuropeptide receptors \cite{beets_system-wide_2022}, combined with gene expression data \cite{taylor_molecular_2021}, provides candidate neuropeptide/GPCR combinations mediating the communication in the majority of neuron pairs we identify (listed in Supporting Spreadsheet 1). For example, M3L and URYVL express the following peptides and receptors, respectively, that can bind with one another: peptide FLP-4 binds to receptor NPR-4 and peptide FLP-5 binds to receptor NPR-11.  Additional peptide/receptor pairs  are also likely expressed, albeit at lower levels, as described in methods.

The screen identifies only those subset of neural connections for which signal propagation is completely absent in \textit{unc-31}. Many more neuron pairs likely signal through multiple parallel paths including both synaptic and  extrasynaptic ones,  or exhibit  co-transmission of both, and these would not appear in the screen. Given the degree of anatomical connectedness (any neuron is anatomically connected to any other in no more than four hops) it is striking that we found so many neuron pairs that pass our stringent test for purely extrasynaptic dependence. 
We note that pairs that pass our screen could include signaling paths that involve   synaptic  signaling \textit{in series} with  extrasynaptic signaling, as this would still be purely extrasynaptic dependent.
Our screen assumes that  the \textit{unc-31} mutant  is defective only for dense core vesicle release, as reported \cite{speese_unc-31_2007}. Defects in any additional signaling modes, or changes to wiring, would present a confound to interpreting these measurements.

We found bilateral partners among the candidate  pairs of neurons identified in our screen for having purely extrasynaptic-dependent connections. These bilateral partners typically had no or only weak wired connections between them, and  anatomy predicted very weak evoked responses (i.e.  below the threshold displayed in Fig.~\ref{fig:anatomy}b and c). AVDR and AVDL is the most prominent bilateral pair found in our screen. Stimulation of AVDR evoked robust \textit{unc-31}-dependent responses in its bilateral partner AVDL. AVDR and AVDL have no or only weak wired connections between them (three of four connectomes show no wired connections, and the fourth finds only a very  weak gap junction). Signal propagation between these bilateral partners is consistent with neuropeptidergic signaling, particularly via autocrine loops, in which a genetically defined cell sub-type expresses both a neuropeptide and the receptor to which it  binds~\cite{ripoll-sanchez_neuropeptidergic_2022, beets_system-wide_2022}.  The AVD cell-type was recently predicted to have a strong autocrine loop based on gene expression and peptide/GPCR interaction studies ~\cite{ripoll-sanchez_neuropeptidergic_2022} mediated by the neuropeptide/GPCR combinations NLP-10->NPR-35 and FLP-6->FRPR-8~\cite{taylor_molecular_2021, beets_system-wide_2022} (Fig.~\ref{fig:esyn}d). 
Moreover, AVD was predicted to be among the top 25  highest-degree ``hub'' nodes in a peptidergic network based on gene expression \cite{ripoll-sanchez_neuropeptidergic_2022}. Consistent with this, we find that AVD is heavily represented among hits in our screen (Extended Data Fig \ref{sfig:esyn}b).

We note that the existence of a neuropeptide/GPCR combination indicates only that the molecular machinery  for extrasynaptic peptidergic signaling is present.  By contrast, functional measurements like those performed here  provide the  more direct evidence that  extrasyanaptic signaling actually occurs. Moreover functional measurements    resolve  whether such signaling  directly  evokes  neural responses, as in this case, or for example, whether they only  modulate neural excitability over  longer-timescales.

\subsection*{Signal-propagation better predicts spontaneous activity than anatomy}
\begin{figure}[htbp]
\centering
\includegraphics[page=1, width=\linewidth]{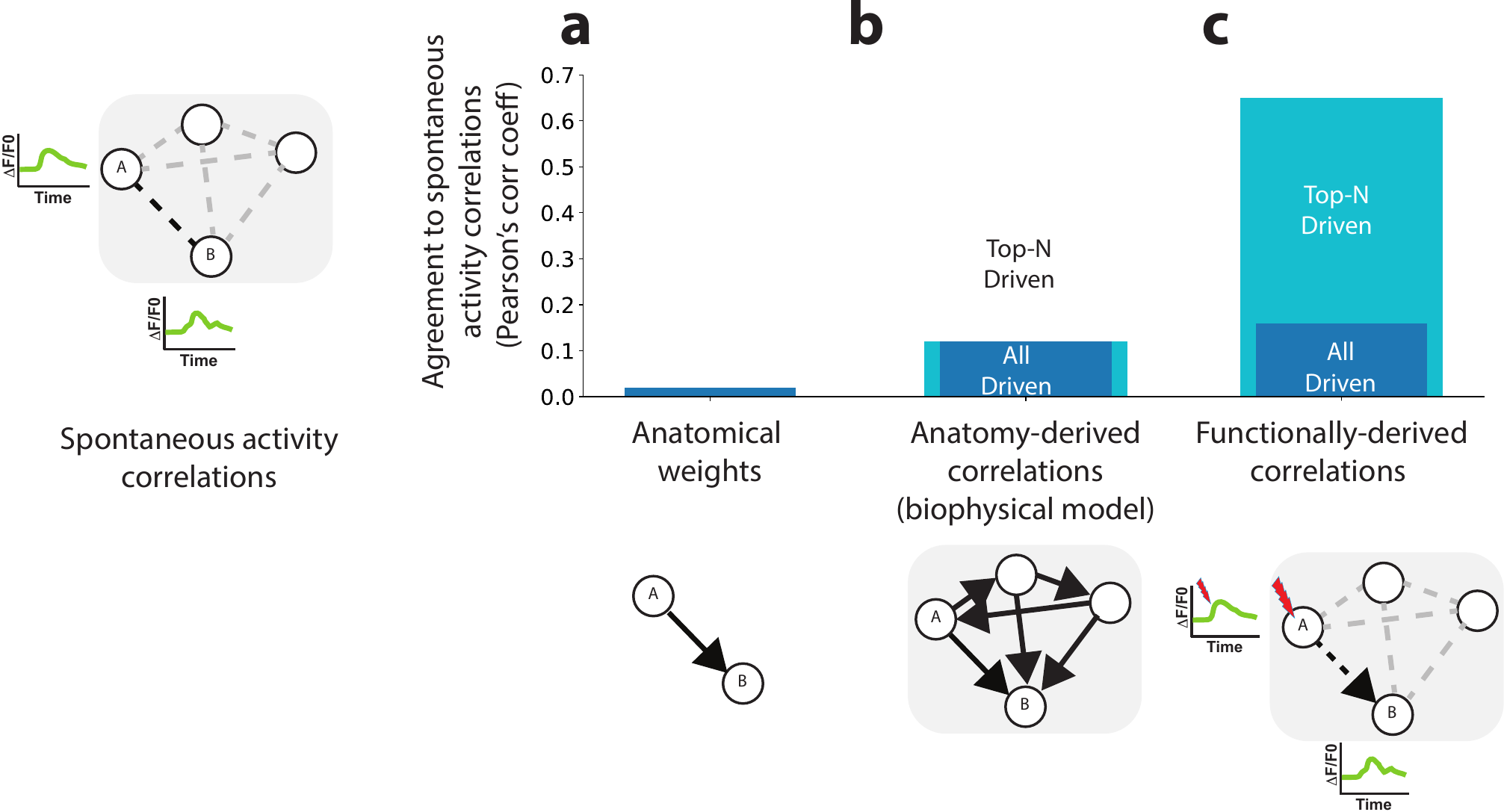}
\caption{\textbf{Measured signal-propagation better predicts spontaneous activity than anatomy.} Schematic illustrates the types of measurements used to predict spontaneous activity correlations. \textbf{a)} Agreement (as Pearson's correlation coefficient) between anatomical weights (synapse counts) and correlation matrix of spontaneous activity recorded from an immobilized animal. \textbf{b)}  Agreement between correlations in anatomy-derived activity and those measured from spontaneous activity. Anatomy-derived activities and correlations are calculated \textit{in silico} by driving all neurons (dark blue) or only an optimal subset of top-n neurons (light-blue). For anatomy, the optimal set is all neurons so the bars are the same. \textbf{c)} Agreement between correlations in  spontaneous activity  and   correlations derived from  functional measurements of signal propagation are reported.  Functionally-derived correlations are calculated  using measured signal-propagation kernels by either: driving all neurons (dark blue) or only an optimal subset, in this case the top 6 (light blue). }  \label{fig:spontaneous}
\end{figure}

A key motivation for mapping neural connections is to understand how they give rise to neural dynamics. We therefore compared how well  functional and anatomical descriptions of the network predict spontaneous neural activity.
We measured the spontaneous network activity of immobilized worms without any optogenetic activators under bright imaging conditions, similar to \cite{kato_global_2015}, and compared correlations in the spontaneous activity to predictions from anatomy and from our  signal propagation measurements (Fig.~\ref{fig:spontaneous}). 
We compared spontaneous activity to two different anatomical descriptions. First, we compared the matrix of  spontaneous activity correlations to the matrix of the bare anatomical weights, or synapse counts between  neurons, (Fig.~\ref{fig:spontaneous}a) which had previously been shown to have poor agreement \cite{yemini_neuropal_2021, uzel_set_2022} and our experiments further  support this conclusion. 
Poor agreement is unsurprising because descriptions  of the monosynaptic contacts only captures  direct connections between two neurons, while activity correlations  are   influenced by all possible paths through the network.  We therefore  also compared spontaneous correlations to anatomical predictions from the connectome-constrained biophysical model  that considers  all anatomical paths through the network. We drove activity in the biophysical model \textit{in silico} and computed correlations from the resulting simulated activity.  These anatomy-derived correlations showed  better but still poor agreement with those measured in spontaneous activity (Fig.~\ref{fig:spontaneous}b).

We also compared spontaneous activity correlations to correlations predicted by our measurements of signal propagation (functionally-derived simulations). The functionally-derived correlations better predicted spontaneous activity than either of the anatomy-based approaches (Fig.~\ref{fig:spontaneous}c).   To derive correlations from our signal propagation measurements we drove neurons  \textit{in silico},  propagated their activity through our measured kernels,  and calculated the correlation matrix from the resulting activity.

We compared performance using two different assumptions: that all neurons drive spontaneous activity (``all driven'') or that only an optimal subset of neurons drive spontaneous activity (``top-$n$ driven''). To find the  subset, we  ranked-ordered each neuron's individual agreement to spontaneous activity and selected the  top-$n$ ranked neurons that collectively had the best agreement when driven. 

Functionally-derived correlations based on  signal propagation measurements  always outperformed the anatomy-based correlations, but the performance is most significant under the assumption that the top-$n$ neurons are driven (Fig.~\ref{fig:spontaneous}c).  Driving activity in the top 6 neurons dramatically improved agreement of the functionally-derived correlations to spontaneous activity compared to the all-driven assumption. Interestingly, for  anatomy-derived correlations, we could not identify a top-n subset of neurons that significantly improved performance (Fig.~\ref{fig:spontaneous}b).  Taken together, we conclude that functionally-derived predictions based on our signal propagation measurements better agree with spontaneous activity correlations than does anatomy, and that some subsets of neurons likely make outsized contributions to driving spontaneous dynamics.

Using our measured signal propagation kernels to simulate the worm's neural activity  
has two distinct advantages compared to the anatomy-derived biophysical model  and all other  previous simulations of \textit{C. elegans}: First, all of the parameters  in the equations governing neural dynamics are extracted  directly from  the measured kernels that describe signal propagation: including the timescales, weights, signs, and connectivity, thereby avoiding ambiguities that arise from interpreting the connectome and transcriptome. Second, unlike all previous models that rely on the connectome, the functional measurements here capture extrasynaptic signaling, such as from neuropeptides, that likely contribute to spontaneous activity but are not visible in anatomy. These two advantages likely both contribute to why functional-derived predictions of correlations outperform anatomy at predicting spontaneous correlations.  An interactive version of our functionally-derived simulation  is  available at \url{https://funsim.princeton.edu}.

\section*{Discussion}
Signal propagation in \textit{C. elegans} measured by neural activation differs from predictions based on anatomy  in part because anatomy fails to account for wireless connections such as extrasynaptic release of neuropeptides \cite{bentley_multilayer_2016}. 
We find that extrasynaptic signaling  serves a functional role  similar to that of classical neurotransmitters by directly evoking calcium activity on seconds timescale. We therefore conclude that extrasynaptic  and  synaptic signals together drive neural dynamics. The role of extrasynaptic signaling  in directly evoking activity is likely in addition to its more well-characterized role in  modulating neural excitability  over  longer-timescales of many minutes.

Our work complements recent efforts to map peptidergic signaling based on gene expression and neuropeptide/GPCR interaction studies \cite{beets_system-wide_2022,ripoll-sanchez_neuropeptidergic_2022}. Gene expression can be used to  identify  pairs of neurons that express the correct  peptides and receptors  for signaling. But measurements of signal propagation, like those performed here, are needed to reveal whether signaling is actually present and to measure the temporal properties and functional role of signaling. In this work we  provided a list of purely extrasynaptic-dependent connections that can be used to validate predictions based on gene-expression.   

Peptidergic extrasynaptic signaling relies on diffusion and therefore \textit{C. elegans'} small size may be uniquely well-suited for this mode of  signal propagation. Mammals  express neuropeptides and receptors, including in multiple brain areas and  throughout mouse cortex \cite{smith_single-cell_2019}, but their larger brain size may limit the speed, strength or spatial extent of peptidergic extrasynaptic signaling.  

Plasticity, neuromodulation, neural network state, and other longer-timescale effects may  contribute to variability in our measured responses or to  discrepancies between anatomical and functional descriptions of the
 \textit{C. elegans} network. A future direction will be to search for  latent connections that may only become functional in certain internal states.

Our \themap{} provides a lower bound on the number of functional connections. Our analysis has fewer observations of neurons whose tagRFP-T expression is too dim or whose color and location pattern is too difficult to identify, and we are limited to calcium activity in the nucleus, and therefore omit compartmentalized calcium dynamics \cite{hendricks_compartmentalized_2012}. %
To better probe nonlinearities in the network, future measurements are needed that explore a larger stimulation space, including simultaneous stimulation of multiple neurons. Future work is also needed to probe functional connections in worms that are unrestrained and free to move.

Our \themap{}   reports  effective connections, not direct connections. Effective connections are  the most relevant and useful for answering the circuit-level questions that motivate our work, such as how a stimulus in one part of the network drives activity in another. By contrast, the properties of direct connections are well suited for probing questions of gene expression, development and anatomy, but can be less informative of network function.  For example,  a direct connection between two neurons may be  slow or weak, but may overlook a fast and strong effective connection via other paths through the network. 
Here we have compared different network descriptions  at the level of effective connections, in this case by using connectome-constrained simulations to derive effective connections from anatomy.
An alternative approach would be to infer properties of direct connections from the measured   effective connections, e.g. following~\cite{randi_nonequilibrium_2021}, but solving this inverse problem  may require a higher  signal-to-noise ratio than our current measurements.

The neural dynamics we observe are slow  but no slower than typical calcium responses to natural stimuli such as odor delivery \cite{lin_functional_2022}. The slow dynamics likely result  from the slow graded potentials in \textit{C. elegans} \cite{narayan_transfer_2011, lindsay_optogenetic_2011},  the slower  calcium dynamics of the  nucleus \cite{kato_global_2015} and  the slow rise and fall time  
 of
GCaMP6s \cite{chen_ultrasensitive_2013} and   
GUR-3/PRDX-2 \cite{bhatla_light_2015}. We identify fast signal transmission, even in  slow dynamics, by fitting kernels  to relate the dynamics of upstream and downstream neurons.  

The signal propagation atlas presented here  informs structure-function investigations at the level of both circuits and whole networks and also enables more accurate brain-wide simulations of neural dynamics and behavior.  Crucially this work quantifies discrepancies between what is expected from anatomy and what is observed functionally at brain scale and cellular resolution for a complete connectome. The finding that extrasynaptic peptidergic signaling directly evokes activity and drives dynamics in \textit{C. elegans} will inform ongoing discussion about efforts to characterize other brains at greater detail and scale.

\bibliographystyle{naturemag-doi}

\section*{Acknowledgments} We thank Annegret Falkner, Mala Murthy, Eva Naumann, H.~Sebastian Seung, Jacob Bien and Josh Shaevitz for comments on the manuscript. Online visualization software was created by Research Computing staff in the Lewis-Sigler Institute for Integrative Genomics and the Princeton Neuroscience Institute with special thanks to Fan Kang, Robert Leach, Ben Singer, and Lance Parsons. 
\section*{Funding} Research reported in this work was supported by the National Institutes of Health  National Institute of Neurological Disorders and Stroke under New Innovator award number DP2-NS116768 to AML; the Simons Foundation under award  SCGB \#543003 to A.M.L.;  by the Swartz Foundation through the Swartz Fellowship for Theoretical Neuroscience to F.R.;  by the National Science Foundation, through the Center for the Physics of Biological Function (PHY-1734030); and by the Boehringer Ingelheim Fonds to S.D.
Strains from this work are being distributed by the CGC, which is funded by the NIH Office of Research Infrastructure Programs (P40 OD010440). 
\section*{Author contributions}
A.M.L.~and F.R.~conceived the experiments,  F.R.~and S.D.~conducted the experiments, A.K.S.~designed and performed all transgenics, F.R.~designed and built the instrument and the analysis framework and pipeline, F.R.~and S.D.~performed  the bulk of the analysis with additional contributions from A.M.L.~and A.K.S. All authors wrote and reviewed the manuscript.
\section*{Data and materials availability}
Data is available in an interactive and browseable format at \url{https://funconn.princeton.edu} and \url{http://funsim.princeton.edu}. Machine readable datasets are publicly accessible through on Open Science Foundation repository at  \url{https://osf.io/e2syt/}. All analysis code is publicly available at \url{https://github.com/leiferlab/pumpprobe}, \url{https://github.com/leiferlab/wormdatamodel}, 
\url{https://github.com/leiferlab/wormneuronsegmentation}, and \url{https://github.com/leiferlab/wormbrain}. Hardware acquisition code is available at \url{https://github.com/leiferlab/pump-probe-acquisition}

\clearpage
\newpage

\section*{Online Methods}

\paragraph*{Worm Maintenance}
\textit{C. elegans} were stored in the dark, and only minimal light was used when transferring worms or mounting worms for experiments. Strains generated in this study (Extended Data Fig.~\ref{ext:strains}a) have been deposited in the Caenorhabditis Genetics Center, University of Minnesota, for public distribution.

\paragraph*{Transgenics}
To measure functional connectivity, we generated the transgenic strain AML462. This strain expresses the calcium indicator GCaMP6s in the nucleus of each neuron, a purple light-sensitive optogenetic protein system (i.e., GUR-3 and PRDX-2) in each neuron, and multiple fluorophores of various colors from the NeuroPAL\cite{yemini_neuropal_2021} system also in the nucleus of neurons. We also used a QF-GR drug-inducible gene expression strategy to turn on  gene expression of optogenetic actuators only later in development. To create this strain, we first generated an intermediate strain, AML456, by injecting a plasmid mix (75 ng/$\mu$l pAS3-5xQUAS::$\Delta$pes-10P::AI::gur-3G::unc-54 + 75 ng/$\mu$l pAS3-5xQUAS::$\Delta$pes-10P::AI::prdx-2G::unc-54 + 35 ng/$\mu$l pAS-3-rab-3P::AI::QF+GR::unc-54 + 100 ng/$\mu$l unc-122::GFP) into CZ20310 worms followed by UV integration and 6 outcrosses \cite{noma_rapid_2018, evans_transformation_2006}. The intermediate strain, AML456, was then crossed into the pan-neuronal GCaMP6s calcium imaging strain, with NeuroPAL, AML320 \cite{yemini_neuropal_2021, yu_fast_2021}. 

An  \textit{unc-31} mutant background with defects in the dense-core vesicle release pathway was used to diminish wireless signaling \cite{speese_unc-31_2007}. We created an \textit{unc-31} knockout version of our functional connectivity strain by performing CRISPR/Cas9-mediated genome editing on AML462 using a single-strand oligodeoxynucleotide (ssODN)-based homology-dependent repair strategy \cite{paix_precision_2017}. This approach resulted in strain AML508 (unc-31 [wtf502] IV; otIs669 [NeuroPAL] V 14x; wtfIs145 [pBX + rab-3::his-24::GCaMP6s::unc-54]; wtfIs348  [75 ng/$\mu$l pAS3-5xQUAS::$\Delta$pes-10P::AI::gur-3G::unc-54 + 75 ng/$\mu$l pAS3-5xQUAS::$\Delta$pes-10P::AI::prdx-2G::unc-54 + 35 ng/$\mu$l pAS-3-rab-3P::QF+GR::unc-54 + 100 ng/$\mu$l unc-122::GFP].

CRISPR/Cas-9 editing was carried out as follows. Protospacer adjacent motif (PAM) sites were selected in the first intron (\texttt{\seqsplit{gagcuucgcaauguugacucCGG}}) and the last intron (\texttt{\seqsplit{augguacauuggguccguggCGG}}) of the \textit{unc-31} gene (ZK897.1a.1) to delete 12,476 out of 13,169 bp (including the 5' and 3' untranslated regions [UTRs]) and 18 out of 20 exons from the genomic locus, while adding 6 bp (GGTACC) for the Kpn-I restriction site (Extended Data Fig.~\ref{ext:strains}b). Alt-R S.p. Cas9 Nuclease V3, Alt-R-single guide RNA (sgRNA), and Alt-R homology-directed repair (HDR)-ODN were used  (IDT, USA). We introduced the Kpn-I restriction site (\texttt{\seqsplit{gacccagcgaagcaaggatattgaaaacataagtacccttgttgttgtgtGGTACCccacggacccaatgtaccatattttacgagaaatttataatgttcagg}}) into our repair oligo to screen and confirm the deletion by PCR followed by restriction digestion. sgRNA and HDR ssODNs were also synthesized for the \textit{dpy-10} gene as a reporter, as described in \cite{paix_precision_2017}. 
An injection mix was prepared by sequentially adding Alt-R S.p. Cas9 Nuclease V3 (1 $\mu$L of 10 $\mu$g/$\mu$L), 
0.25 $\mu$L of 1M KCL, 0.375 $\mu$L of 200 mM HEPES (pH 7.4), 
sgRNAs for \textit{unc-31} [1 $\mu$L each for both sites], and 
0.75 $\mu$L for \textit{dpy-10} from a stock of 100 $\mu$M, 
ssODNs [1 $\mu$L for \textit{unc-31} and 0.5 $\mu$L for \textit{dpy-10} 
from a stock of 25 $\mu$M], and nuclease-free water to a final volume of 10 $\mu$L in a PCR tube, kept on ice. 
The injection mix was then incubated at 37 \degree C for 15 min before it was injected into the germline of AML462 worms. 
Progenies from plates showing roller or dumpy phenotypes in the F1 generation post-injection were individually propagated and PCR>Kpn-I digestion screened to confirm deletion. Single-worm PCR was carried out using GXL-PRIME STAR taq-Polymerase (Takara Bio, USA) and the Kpn-1-HF restriction enzyme (NEB, USA). Worms without a roller or dumpy phenotype and homozygous for deletion were confirmed by Sanger sequencing fragment analysis.

To cross validate GUR-3/PRDX-2 evoked behavior responses, we generated the transgenic strain AML546 by injecting a plasmid mix (40 ng/$\mu$l pAS3-rig-3P::AI::gur-3G::SL2::tagRFP::unc-54  + 40 ng/$\mu$l pAS3-rig-3P::AI::prdx-2G::SL2::tagBFP::unc-54) into N2- worms to generate transient transgenic line expressing GUR-3/PRDX-2 in AVA neurons. 

\paragraph*{Cross-validation of GUR-3/PRDX-2 evoked behavior}
Optogenetic activation of AVA neurons using traditional channelrhodopsins (e.g. Chrimson) leads to reversals~\cite{gordus_feedback_2015,li_encoding_2014}. We used worms expressing GUR-3/PRDX-2 in AVA neurons (AML 564) to show that GUR-3/PRDX-2 elicits a similar behavioral response. We illuminated freely moving worms with blue light from an LED (peaked at 480 nm, $2.3 mW/mm^2$) for 45 s. We compared the number of onsets of reversals in that period of time with a control in which only dim white light was present, as well with the results of the same assay performed on N2 worms. Animals with GUR-3/PRDX-2 in AVA (n=11 animals) exhibited more blue-light evoked reversals per minute than WT (n=8 animals) (Extended Data Fig.~\ref{ext:stim_characterization}h).

\paragraph*{Dexamethasone treatment}
To increase expression of optogenetic proteins  while avoiding toxicity during the animals' early life development, a drug-inducible gene expression strategy was used. Dexamethasone  (dex) %
activates QF-GR to temporally control the expression of downstream targets \cite{monsalve_new_2019}, in this case the optogenetic proteins  in the functional connectivity imaging strains AML462 and AML508. 
Dex-NGM plates were prepared by adding  200 $\mu$M of dex in DMSO just before pouring the plate. For dex treatment, L2/L3 worms were transferred to overnight-seeded dex-NGM plates and further grown until worms were ready for imaging. Further details of the dex treatment are described below.

We prepared stock solution of 100 mM dex  by dissolving 1 gram Dexamethasone (D1756, Sigma-Aldrich) in 25.5 ml DMSO (D8418, Sigma-Aldrich). Stocks were then filter sterilized, aliquoted, wrapped in foil to prevent light, and stored at  -80 C until needed.   200 $\mu$M dex-NGM plates were made by adding 2 ml of 100 mM dex stock in 1 liter of NGM-agar media, while stirring, 5 minutes before pouring the plate. Dex-plates were stored at 4 C for up to a month  until needed.

\paragraph*{Preparation of worms for imaging}
Worms were individually mounted on 10 \% agarose pads prepared with M9 buffer and  immobilized using 2 $\mu$l of 100 nm polystyrene beads solution and 2 $\mu$l of levamisole (500 $\mu$M stock). This concentration of levamisole, after dilution in the polystyrene bead solution and the agarose pad water, largely immobilized the worm while still allowing the worm to slightly move, especially before placing the coverslip. Pharyngeal pumping was  observed during imaging.

\paragraph*{Multi-channel imaging and neural identification}
Volumetric, multi-channel imaging was performed to capture  images of the following fluorophores in the  Neuropal transgene: mtagBFP2, CyOFP1.5, tagRFP-T, and mNeptune2.5 \cite{yemini_neuropal_2021}. Light downstream of the same spinning disk unit used for calcium imaging  traveled on an alternative light path through channel-specific filters mounted on a mechanical filter wheel, while mechanical shutters alternated illumination with the respective lasers, similar to \cite{yu_fast_2021}. Channels were as follows:  mtagBFP2 was imaged using a 405 nm laser and a Semrock FF01-440/40 emission filter; CyOFP1.5 was imaged using a 505 nm laser and a Semrock 609/54 emission filter; tagRFP-T was imaged using a 561 nm laser and a Semrock 609/54 nm emission filter; and mNeptune2.5 was imaged using a 561 nm laser and a Semrock 732/68 nm emission filter. 

After the functional connectivity recording was complete, a human manually assigned neuron identities by comparing each neuron's color, position, and size to a known atlas. Some neurons are particularly hard to identify in NeuroPAL and are therefore absent or less frequently identified in our recordings. Some neurons have dim tagRFP-T expression, which makes it difficult for the neuron segmentation algorithm to find them and, therefore, to extract their calcium activity. These neurons include, for example, AVB, ADF, and RID. RID's distinctive position and its expression of CyOFP allowed us nevertheless to manually target it optogenetically. Neurons in the ventral ganglion are hard to identify because it appears as very crowded when viewed in the most common orientation that worms assume when mounted on a microscope slide. Neurons in the ventral ganglion are therefore sometimes difficult to distinguish one from another, especially for dimmer neurons such as the  SIA, SIB, and RMF neurons. In our strain, the neurons  AWCon and AWCoff were  difficult to tell apart based on color information.

\paragraph*{Volumetric image acquisition}

Neural activity was recorded at whole-brain scale and cellular resolution via continuous acquisition of volumetric images in red and green channels with a spinning disk confocal unit and via LabView software (\url{https://github.com/leiferlab/pump-probe-acquisition/tree/pp}), similarly to~\cite{nguyen_whole-brain_2015}, with a few upgrades. The imaging focal plane was scanned through the brain of the worm remotely using  an electrically tunable lens (Optotune EL-16-40-TC) instead of moving the objective. The use of remote focusing allowed us to decouple the z-position of the imaging focal plane and that of the optogenetics 2-photon spot (described below). 

Images were acquired by an sCMOS camera, and each acquired image frame was associated to the focal length of the tunable lens (z-position in the sample) at which it was acquired. To ensure the correct association between frames and z-position, we recorded the analog signal describing the focal length of the tunable lens at time points synchronous with a trigger pulse output by the camera. By counting the camera triggers from the start of the recording, the z-positions could be associated to the correct frame, bypassing unknown operating-system-mediated latencies between the image stream from the camera and acquisition of analog signals.

Additionally,  real-time ``pseudo''-segmentation of the neurons (described below)  required the ability to separate frames into corresponding volumetric images in real-time. Because the z-position was acquired at low sample rate, splitting of volumes based on finite differences between successive z-positions could lead to errors in  assignment at the edge of the z-scan. An analog OP-AMP-based differentiator was used to independently detect the direction of the z-scan in hardware.

\paragraph*{Calcium imaging}
Calcium imaging was performed in single-photon regime with a 505 nm excitation laser via spinning disk confocal microscopy, at 2 vol/s. For functional connectivity experiments, an intensity of 1.4 mW/mm$^2$ at the sample plane was used to image GCaMP6s, well below the threshold needed to excite the GUR-3/PRDX-2 optogenetic system \cite{bhatla_light_2015}. We note that at this wavelength and intensity animals exhibited very little spontaneous calcium activity.

For certain analyses  (Fig.~\ref{fig:spontaneous}), recordings with ample spontaneous activity were desired.  In those cases, we  increased the 505 nm intensity seven-fold to approximately 10 mW/mm$^2$ 
and recorded  from AML320 strains that lacked exogenous GUR-3/PRDX-2 to avoid potential widespread neural  activation. Under these imaging conditions, we observed  population-wide slow stereotyped spontaneous oscillatory calcium dynamics, as previously reported \cite{kato_global_2015, hallinen_decoding_2021}.

\paragraph*{Extraction of calcium activity from the images}
The extraction of calcium activity from the raw images was performed using Python libraries implementing optimized versions of the algorithm described in~\cite{nguyen_automatically_2017}, available at \url{https://www.github.com/leiferlab/pumpprobe},\\ \url{https://www.github.com/leiferlab/wormdatamodel},\\ \url{https://www.github.com/leiferlab/wormneuronsegmentation-c},\\ and \url{https://www.github.com/leiferlab/wormbrain}. 

The positions of neurons in each acquired volume was determined by  computer vision software implemented in C++.  This software was greatly optimized to identify neurons in real-time in order to also enable closed-loop targeting and stimulus delivery (as described in the section \textbf{Stimulus delivery and pulsed laser}). Two design choices made this algorithm dramatically faster than previous approaches.
First, a  local maxima search was used instead of a slower  watershed-type segmentation. The nuclei of \textit{C. elegans} neurons are approximately spheres and so they can be identified and separated by a simple local maxima search. 
Second, we factorized the 3D local maxima search into multiple 2D local maxima searches. In fact, any local maximum in a 3D image is also a local maximum in the 2D image in which it is located. Local maxima were therefore first found in each 2D image separately, and then candidate local maxima were discarded or retained by comparing them to their immediate surroundings in the other planes. This makes the algorithm less computationally intensive and fast enough to be used also in real time. 
We refer to this type of algorithm as ``pseudo''-segmentation because it finds the center of neurons without fully describing the extent and boundaries of each neuron.

After neural locations were found in each of the volumetric images, a nonrigid pointset registration algorithm was used to track their locations across time, matching neurons identified in a given 3D image to the neurons identified in a 3D image chosen as reference.  Even worms that are mechanically immobilized  still  move slightly and contract their pharynx, thereby deforming their brain and requiring  the tracking of neurons. We implemented in C++ a fast and optimized version of the Dirichelet-Student-t Mixture Model (DSMM) ~\cite{zhou_accurate_2018}.

\paragraph*{Calcium pre-processing}
The GCaMP6s intensity extracted from the images undergoes the following pre-processing steps. (1) Missing values are  interpolated based on neighboring time points. Missing values can occur when a neuron cannot be identified in a given volumetric image. (2) Photobleaching is removed by fitting a double exponential to the baseline signal. (3) Outliers more than 5 standard deviations away from the average are removed from each trace. (4) Traces are smoothed via a causal polynomial filtering with a window size of 6.5 s and polynomial order of 1 [Savitzky--Golay filters with windows completely ``in the past'', e.g., obtained with \texttt{scipy.signal.savgol\_coeffs(window\_length=13,\\ polyorder=1, pos=12)}]. This type of filter with the chosen parameters is able to remove noise without smearing the traces in time. Note that when fits are  performed (e.g., to calculate kernels), they are always performed on the original, non-smoothed traces. (5) Where $\Delta F/F_0$ of responses is used, $F_0$ is defined as the value of $F$ in a short interval before the stimulation time.

\paragraph*{Stimulus delivery and pulsed laser}
For two-photon optogenetic targeting, we used an optical parametric amplifier (OPA,  Light Conversion ORPHEUS) pumped by a femtosecond amplified laser (Light Conversion PHAROS). The output of the OPA was tuned to a wavelength of 850 nm, at a 500 kHz repetition rate. We used temporal focusing to spatially restrict the size of the 2-photon excitation spot along the microscope axis. A motorized iris was used to set its lateral size. For temporal focusing, the first-order diffraction from a reflective grating, oriented orthogonally to the microscope axis, was collected (as in \cite{papagiakoumou_patterned_2008}) and traveled through the motorized iris, placed on a plane conjugate to the grating. To arbitrarily position the 2-photon excitation spot in the sample volume, the beam then traveled through an electrically tunable lens (Optotune EL-16-40-TC, on a plane conjugate to the objective), to set its position along the microscope axis, and finally was reflected by two galvo-mirrors to set its lateral position. The pulsed beam was then combined with the imaging light path by a dichroic mirror immediately before entering the back of the objective.

The majority of the stimuli were delivered automatically by computer control. Real-time computer vision software found the position of the neurons for each volumetric image acquired, using only the tagRFP-T channel. To find neural positions, we used the same ``pseudo''-segmentation algorithm described above. The algorithm found neurons in each 2D frame in $\sim $500 $\mu$s as the frames arrived from the camera. In this way locations for all neurons in a volume were found within a few milliseconds of acquiring the last frame of that volume.

Every 30 s, a random neuron was selected among the neurons found in the current volumetric image based only on its tag-RFP-T signal. After galvo-mirrors and tunable lens set the position of the 2-photon spot on that neuron, a 500 ms (300 ms for the \textit{unc-31} mutant strain) train of light pulses was used to optogenetically stimulate that neuron. 
The duration of stimulus illumination for the \textit{unc-31} mutant strain was selected  to elicit calcium transients in stimulated neurons with a distribution of amplitudes such that the maximum amplitude was similar to those in WT background animals, (Extended Data Fig.~\ref{sfig:autoresponse_amplitude}f). 
The output of the laser was controlled via the external interface to its built-in pulse picker, and the power of the laser  at the sample was 1.2 mW at 500 kHz. Neuron identities were assigned to stimulated neurons after the completion of experiments using NeuroPAL \cite{yemini_neuropal_2021}.

To probe the AFD-AIY neural connection, a small set of stimuli used variable pulse durations from 100 ms to 500 ms in steps of 50 ms selected randomly to vary the amount of optogenetic activation of AFD.

In some cases, neurons of interest were too dim to be detected by the real-time software. For those neurons of interest, additional recordings were performed in which a human manually selected the neuron to be stimulated  based on its color, size, and position. This was the case for certain stimulations of neurons RID and AFD.

\paragraph*{Characterization of the 2-photon excitation spot size}
The lateral (xy) size of the 2-photon excitation spot was measured with a fluorescent microscope slide, while the axial (z) size was measured using 0.2 nm fluorescent beads (Suncoast Yellow, Bangs Laboratories), by scanning the z-position of the optogenetic spot while maintaining the imaging focal plane fixed (Extended Data Fig.~\ref{sfig:spot_wavelength}a).

We further tested our targeted stimulation in two ways: selective photobleaching and neuronal activation. First, we targeted individual neurons at various depth in the worm's brain, and we illuminated them with the pulsed laser to induce selective photobleaching of tagRFP-T. Extended Data Fig.~\ref{sfig:bleaching}c,d shows how our 2-photon excitation spot selectively targets individual neurons, because it photobleaches tagRFP-T only in the neuron that we decide to target, and not in nearby neurons. To faithfully characterize the spot size, we set the laser power such that the 2-photon interaction probability profile of the excitation spot would not saturate the 2-photon absorption probability of tagRFP-T. Second, we showed that our excitation spot is restricted along the z-axis by targeting a neuron and observing its calcium activity. When the excitation was directed at the neuron but shifted by 4 $\mu$m along z, the neuron showed no activation. In contrast, the neuron showed activation when the spot was correctly positioned on the neuron (Extended Data Fig.~\ref{sfig:walkoff}e).

\paragraph*{Inclusion criteria}
Stimulation events were included for further analysis if they evoked a detectable calcium response in the stimulated neuron. A classifier determined whether the response was detected by inspecting whether the amplitude of the $\Delta F/F_0$ transient and its second derivative exceeded hand-tuned thresholds. Stimulation events that did not meet this threshold were excluded. RID responses shown in Fig.~\ref{fig:wireless} and Extended Data Fig.~\ref{sfig:RID_more_responses}c are an exception to this policy. RID is visible  based on its CyOFP expression, but its tagRFP-T  expression is too dim to consistently extract calcium signals. Therefore in Fig.~\ref{fig:wireless} and Extended Data Fig.~\ref{sfig:RID_more_responses}c (but not in other figures, like Fig.~\ref{fig:map}) responses to  RID stimulation were included even in cases where it was not possible to extract a calcium-activity trace in RID.

Neuron traces were excluded from analysis if a human was unable to assign an identity or if the imaging time points were absent in a contiguous segment longer than 5\% of the response window due to imaging artifacts or tracking errors. A different policy applies to dim neurons of interest that are not automatically detected by the ``pseudo''-segmentation algorithm in the 3D image used as reference for the pointset registration algorithm. In those cases, we manually added the position of those neurons to the reference 3D image. If these ``added'' neurons are automatically detected in most of the other 3D images, then a calcium activity trace can be successfully produced by the DSMM nonrigid registration algorithm and is treated as any other trace. However, if the ``added'' neurons are too dim to be detected also in the other 3D images and the calcium activity trace cannot be formed for more than 50\% of the total time points, the activity trace for those neurons is extracted from the neuron's position as determined from the position of neighboring neurons. In the analysis code, we refer to these as ``matchless'' traces, because the reference neuron is not matched to any detected neuron in the specific 3D image, but its position is just transformed according to the DSMM nonrigid deformation field. In this way, we are able to recover the calcium activity also of some neurons whose tag-RFP-T expression is otherwise too dim to be reliably detected by the ``pseudo''-segmentation algorithm. Responses to RID stimulation shown in Fig.~\ref{fig:wireless} and Extended Data Fig.~\ref{sfig:RID_more_responses}c are an exception to this policy.  There, the activity of any neuron for which there is not a trace for more than 50\% of the time points is substituted with the corresponding ``matchless'' trace, and not just for the manually added neurons. This is important to be able to show responses of neurons like ADL, which have dim tagRFP-T expression. In this RID-specific case, in order to exclude responses that become very large solely because of numerical issues in the division by the baseline activity due to the dim tagRFP-T, we additionally introduce a threshold at $\Delta F/F=2$.

Kernels were computed only for stimulation-response events for which the automatic classifier detected responses in \textit{both} the stimulated  and downstream neurons. If the downstream neuron did not show a response,  we considered the downstream response to be below the noise level and the kernel to be zero. 

\paragraph*{Statistical analysis}
To assess the relative significance of a  functional connection between a target and putative responding neuron, we calculated the probability of observing  the measured calcium response  given no neural stimulation.  
We used a two-sided Kolmogorov--Smirnov test to compare  the distributions of $\Delta F/F_0$ and its temporal second derivative from  all observations of that neuron pair to the empirical null distributions  from  control recordings lacking stimulation.
$p$-values were calculated separately for   $\Delta F/F_0$ and its temporal second derivative, and then combined  using Fischer's method to report a single fused $p$-value for each neuron pair.  
Finally, to account for the large number of hypotheses tested, a false discovery rate was estimated. From the list of $p$-values,  each neuron was assigned a $q$-value via the Storey--Tibshirani method \cite{storey_statistical_2003}.  $q$-values are interpreted as follows: when considering an ensemble of putative functional connections  of $q$-values all less than or equal to $q_c$,  approximately $q_c$ of those connections would have appeared in a recording that lacked any stimulation. 

To explicitly test whether a pair of neurons are functionally not-connected, taking into account the amplitude of the response, their reliability, and the number of observations, we also computed equivalence $p_{\mathrm{eq}}$ and $q_{\mathrm{eq}}$-values. This asseses the confidence of a pair \emph{not} being connected. We test whether our response is equivalent to what we would expect from our control distribution using the two one-sided t-tests (TOST)~\cite{schuirmann_comparison_1987}. We computed $p_{\mathrm{eq}}$-values for $\Delta F/F_0$ and its temporal second derivative for a given pair being equivalent to the control distributions within an $\epsilon=1.2\sigma_{\Delta F/F_0,\partial_t^2}$. Here, $\sigma_{\Delta F/F_0,\partial_t^2}$ is the standard deviation of the corresponding control distribution. We then combined the two $p_{\mathrm{eq}}$-values into a single one with the Fisher method and computed $q_{\mathrm{eq}}$-values. Note that, differently from the regular $p$-values described above, the equivalence test relies on the arbitrary choice of $\epsilon$, which defines when we call two distributions equivalent. We chose a conservative value of $\epsilon=1.2\sigma$.

\paragraph*{Measuring path length through the synaptic network}
To find the minimum path length between neurons in the anatomical network topology, we proceeded iteratively. We started from the original binary connectome and computed the map of strictly 2-hop connections by looking for pairs of neurons that are not connected in the starting connectome (the actual anatomical connectome at the first step) but that are connected through a single intermediate neuron. To generate the strictly 3-hop connectome,  we repeated this procedure using the binary connectome including direct and 2-hop connections, as the starting connectome. This process continued iteratively to generate the strictly n-hop connectome.

In the anatomical connectome (the starting connectome for the first step in the procedure above), a neuron was considered to be directly anatomically connected if the connectomes of any of the four L4 or adult individuals in \cite{white_structure_1976} and \cite{witvliet_connectomes_2021} contained at least one synaptic contact between them. Note that this is a permissive description of anatomical connections as it considers even neurons with only a single synaptic contact in only one individual to be connected.

\paragraph*{Fitting kernels}
Kernels $k_{ij}(t)$ were defined as the functions to be convolved with the activity $\Delta F_j$ of the stimulated neuron to obtain the activity $\Delta F_i$ of a responding neuron $i$, such that $\Delta F_i(t) = \big(k_{ij}*\Delta F_j\big)(t)$. To fit kernels, each kernel $k(t)$ was parametrized as a sum of convolutions of decaying exponentials
\begin{equation}
    \label{eq:parametrization}
    k(t) = \sum\limits_m c_m \big(\theta(t) e^{-\gamma_{m,0}t}\big)*\big(\theta(t) e^{-\gamma_{m,1}t}\big)*...,
\end{equation}
where the indices $i,j$ are omitted for clarity and $\theta$ is the Heaviside function. This parametrization is exact for linear systems, and works as a description of causal signal transmission also in nonlinear systems. Note that increasing the number of terms in the successive convolutions does not lead to overfitting, as would occur by increasing the degree of a polynomial. Overfitting could occur by increasing the number of terms in the sum, which in our fitting is constrained to be a maximum of 2. The presence of two terms in the sum allows the kernels to represent signal transmission with saturation (with $c_0$ and $c_1$ of opposite signs) and assume a fractional-derivative-like shape.

The convolutions are performed symbolically. The construction of kernels as in Eq.~\eqref{eq:parametrization} starts from a symbolically stored, normalized decaying exponential kernel with a factor $A$, $A\gamma_0\theta(t)e^{-\gamma_0 t}$. Convolutions with normalized exponentials $\gamma_n \theta(t) e^{-\gamma_n t}$ are performed sequentially and symbolically, taking advantage of the fact that successive convolutions of exponentials always produce a sum of functions in the form $\propto\theta(t)t^n e^{-\gamma t}$. Once rules are found to convolve an additional exponential with a function in that form, any number of successive convolution can be performed. These rules are as follows:
\begin{enumerate}
        \item If the initial term is a simple exponential with a given factor (not necessarily just the normalization $\gamma$) $c_i\theta(t) e^{-\gamma_i t}$ and $\gamma_i\neq\gamma_n$,  then the convolution is
        \begin{equation}
        c_i\theta(t) e^{-\gamma_i t} * \gamma_n \theta(t) e^{-\gamma_n t} = c_\mu\theta(t)e^{-\gamma_\mu t} + c_\nu\theta(t) e^{-\gamma_\nu t},
        \end{equation}
        with $c_\mu=\frac{c_i\gamma_n}{\gamma_n-\gamma_i}$, $c_\nu = - \frac{c_i\gamma_n}{\gamma_n-\gamma_i}$, and $\gamma_\mu=\gamma_i$, $\gamma_\nu=\gamma_n$.
        
        \item If the initial term is a simple exponential and $\gamma_i = \gamma_n$, then
        \begin{equation}
        c_i\theta(t) e^{-\gamma_i t} * \gamma_n \theta(t) e^{-\gamma_n t} = c_\mu\theta(t) t e^{-\gamma_\mu t},
        \end{equation}
        with $c_\mu = c_i \gamma_i$ and $\gamma_\mu=\gamma_i$.
        
        \item If the initial term is a $c_i \theta(t) t^n e^{-\gamma_i t}$ term and $\gamma_i=\gamma_\mu$, then
        \begin{equation}
        c_i\theta(t) t^n e^{-\gamma_i t} * \gamma_n \theta(t) e^{-\gamma_n t} = c_\mu \theta(t) t^{n+1} e^{-\gamma_\mu t},
        \end{equation}
        with $c_\mu = \frac{c_i\gamma_i}{n+1}$ and $\gamma_\mu = \gamma_i$.
        
        \item If the initial term is a $c_i \theta(t) t^n e^{-\gamma_i t}$ term and $\gamma_i\neq\gamma_\mu$, then
        \begin{equation}
        c_i\theta(t) t^n e^{-\gamma_i t} * \gamma_n \theta(t) e^{-\gamma_n t} = c_\mu\theta(t)t^ne^{-\gamma_\mu t} + c_\nu \big( \theta(t) t^{n-1} e^{-\gamma_i t} * \theta(t) e^{-\gamma_n t} \big),
        \end{equation}
        where $c_\mu = \frac{c_i\gamma_n}{\gamma_n-\gamma_i}$,  $\gamma_\mu = \gamma_i$, and $c_\nu = -n\frac{c_i\gamma_n}{\gamma_n-\gamma_i}$.
\end{enumerate}

Additional terms in the sum in Eq.~\eqref{eq:parametrization} can be introduced by keeping track of the index $m$ of the summation for every term and selectively convolving new exponentials only with the corresponding terms.

\paragraph*{Kernel-based simulations of activity}
Using the kernels fitted from our functional data, we can simulate neural activity without making any further assumptions about the dynamical equations of the network of neurons. To compute the response of a neuron $i$ to the stimulation of a neuron $j$, we simply convolve the kernel $k_{i,j}(t)$ with the activity $\Delta F_j(t)$ induced by the stimulation in neuron $j$. The activity of the stimulated neuron can be either the experimentally observed activity or an arbitrarily shaped activity introduced for the purposes of simulation.

To compute kernel-derived neural activity correlations (Fig.~\ref{fig:spontaneous}), we completed the following steps.
(1) We computed the responses of all the neurons $i$ to the stimulation of a neuron $j$ chosen to drive activity in the network. To compute the responses, for each pair $i,j$, we used the kernel $\langle k_{i,j}(t)\rangle_\mathrm{trials}$ averaged over multiple trials. For kernel-based analysis, pairs with connections of $q>0.05$ were considered not connected. We   set the activity  $\Delta F_j(t)$ in the driving neuron to mimic  an empirically observed representative activity transient.
(2) We computed the correlation coefficient of the resulting activities.
(3) We repeated steps 1 and 2 for a set of driving neurons (all or top n neurons, as in Fig.~\ref{fig:spontaneous}).
(4) For each pair $k,l$, we took the average of the correlations obtained by driving the set of neurons $j$ in step 3.

\paragraph*{Anatomy-derived simulations of activity}
Anatomy-derived simulations were performed as described in \cite{kunert_low-dimensional_2014}. Briefly this simulation approach  uses differential equations to model signal transmission via electrical and chemical synapses and includes a nonlinear equation for synaptic activation variables. We injected current \textit{in silico} into individual neurons and simulated the responses of all the other neurons. Anatomy-derived responses (Fig.~\ref{fig:anatomy}) of the connection from neuron $j$ to neuron $i$ were computed as the peak of the response of neuron $i$ to the stimulation of $j$. Anatomy-based predictions of spontaneous correlations were calculated analogously to kernel-based predictions. 

In one analysis in Fig.~\ref{fig:anatomy}d the synapse weights and polarities were allowed to float and were fitted from the functional measurements.  In all other cases, synapse  weights were taken as the scaled average of three adult connectomes from \cite{white_structure_1986, witvliet_connectomes_2021} and an L4 connectome from \cite{witvliet_connectomes_2021} and  polarities were assigned based on a gene expression analysis of ligand-gated ionotropic synaptic connections that considered glutamate, acetylcholine, and GABA neurotransmitter and receptor expression as performed in  \cite{fenyves_synaptic_2020} and  taken from CeNGeN \cite{taylor_molecular_2021} and other sources. Specifically, we used a previously published dataset \url{https://doi.org/10.1371/journal.pcbi.1007974.s003} and aggregated polarities across all members of a cellular subtype (e.g.,~ polarities from source AVAL and AVAR were combined).  In cases of ambiguous polarities, connections were assumed to be excitatory as in \cite{fenyves_synaptic_2020}.  For other biophysical parameters we chose values commonly used in  \textit{C. elegans} modeling efforts \cite{wicks_dynamic_1996, kunert_low-dimensional_2014, kunert-graf_multistability_2017, izquierdo_head_2018}.

\paragraph*{Characterizing stereotypy of functional connections}
To characterize the stereotypy of a neuron pair's functional connection, its kernels were inspected.  A kernel was calculated for every stimulus-response event in which both the upstream and downstream neuron exhibited activity that exceeded a threshold. At least two stimulus-response events that exceeded this threshold were required in order to calculate their stereotypy. The general strategy for calculating stereotopy was to convolve  different kernels with the same stimulus inputs and compare the resulting outputs. Similarity of two outputs is reported as a Pearson's correlation coefficient. Kernels corresponding to different stimulus-response events of the same pair of neurons were compared with one another round-robin style, one round-robin each for a given input stimuli. For inputs we chose the set of all stimuli delivered to the upstream neuron. The neuron-pairs stereotypy is reported as the average Pearsons's correlation coefficient across all round-robin kernel pairings and across all stimuli.

\paragraph*{Rise-time of kernels}
The rise time of kernels, displayed in Fig.~\ref{fig:esyn}c and Experimental Data Fig.~\ref{sfig:kernels}d, was defined as the interval between the earliest time at which the value of the kernel was $1/e$ its peak value and the  time of its peak (whether positive or negative). The rise time was zero if the peak of the kernel was at time t=0. 
However, saturation of the signal transmission can make kernels appear slower than the connection actually is. For example, the simplest instantaneous connection would be represented by a single decaying exponential in Eq.~\ref{eq:parametrization}, which would have its peak at time $t=0$. However, if that connection is saturating, a second, opposite-sign term in the sum is needed to fit the kernel. This second term would make the kernel have a later peak, thereby masking the instantaneous nature of the connection. To account for this effect of saturation, we removed terms representing saturation from the kernels and found the rise time of these ``non-saturating'' kernels.

\paragraph*{Screen for purely extrasynaptic-dependent connections}
To find candidate purely extrasynaptic-dependent connections we considered the pairs of neurons that are connected in WT animals ($q^\mathrm{WT}<0.05$) and non-connected in unc-31 animals ($q_\mathrm{eq}^{unc-31}$<0.05, with the additional condition $q^{unc-31}>0.05$ to exclude very small responses that are nonetheless significantly different from the control distribution.)
We list these connections and provide additional examples in  Extended Data Fig ~\ref{stab:esyn}. 

Using recent neuropeptide/GPCR interaction screen in \textit{C. elegans}~\cite{beets_system-wide_2022} and gene expression data from CeNGEN~\cite{taylor_molecular_2021}, we find putative combinations of neuropeptide and GPCR that can mediate those connections ( Supporting Spreadsheet 1). We produced such list of neuropeptide and GPCR combinations using the Python package Worm Neuro Atlas \href{https://github.com/francescorandi/wormneuroatlas}{https://github.com/francescorandi/wormneuroatlas}. In the list, we only include transcripts from CeNGEN detected with the highest confidence (threshold 4), following~\cite{ripoll-sanchez_neuropeptidergic_2022}. For each neuron pair, we first searched the CeNGEN database for neuropeptides expressed in the upstream neuron, then identified potential GPCR targets for each neuropeptide using information from~\cite{frooninckx_neuropeptide_2012,beets_system-wide_2022}, and finally went back to the CeNGEN database to find if the downstream neuron in the pair was among the neurons expressing the specific GPCRs. The existence of potential combinations of neuropeptide and GPCR putatively mediating signaling supports our observation that communication in the candidate neuron pairs we identify can indeed be mediated extrasynaptically via neuropeptidergic machinery.

\newpage
\renewcommand{\figurename}{Extended Data Figure}
\renewcommand{\tablename}{Extended Data Table}
\renewcommand{\thefigure}{\arabic{figure}}
\renewcommand{\thetable}{\arabic{table}}
\setcounter{figure}{0}
\setcounter{table}{0}

\section*{Extended Data}

\begin{figure}[htbp]
\centering
\includegraphics[page=1, width=\linewidth]{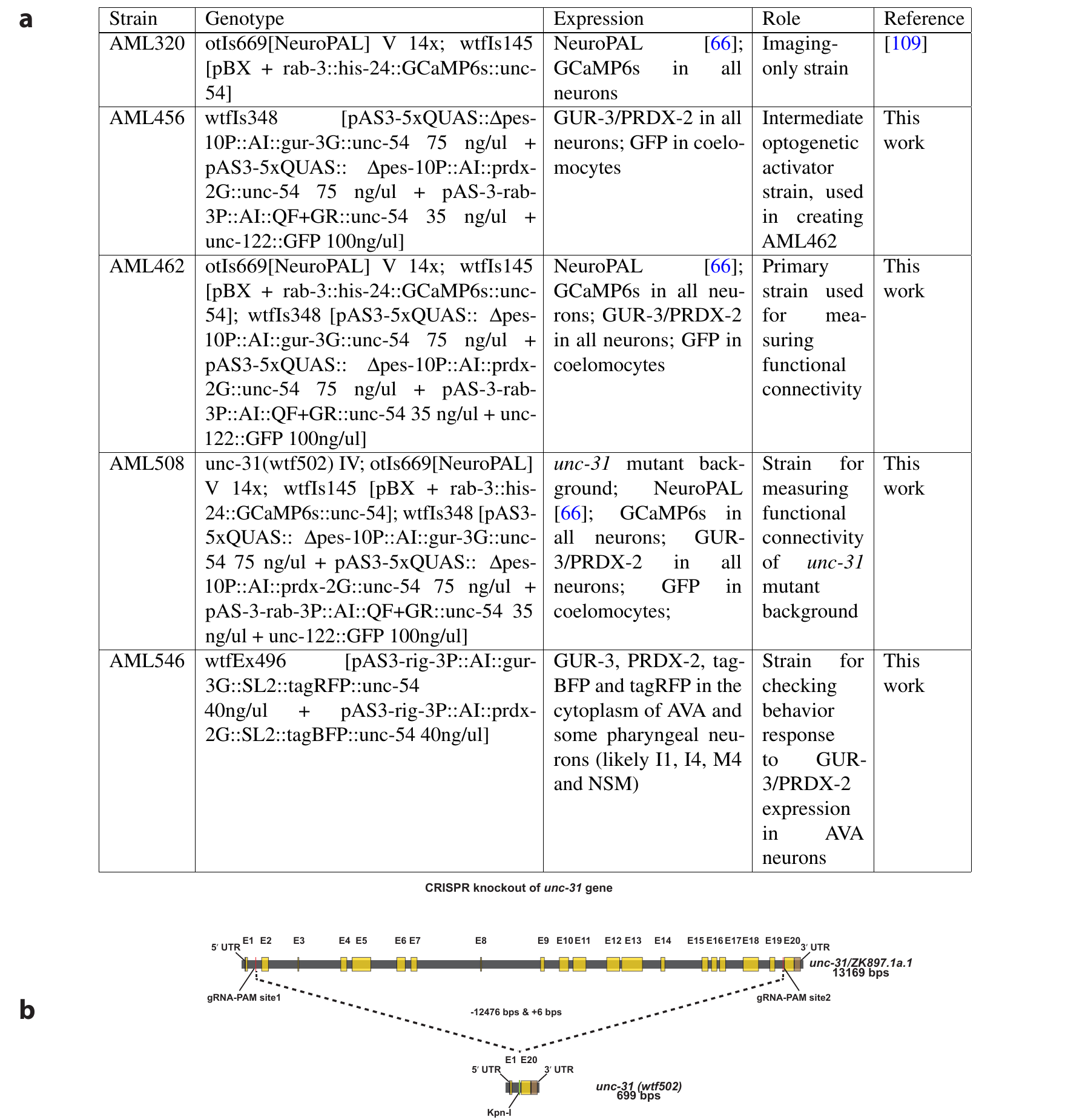}
\caption{ \textbf{Strains a)} Table of strains used in this work. \textbf {b)}  Schematic of CRISPR knockout of \textit{unc-31}.  }  \label{ext:strains}
\end{figure}

\begin{figure}[htbp]
\centering
\includegraphics[page=2, width=\linewidth]{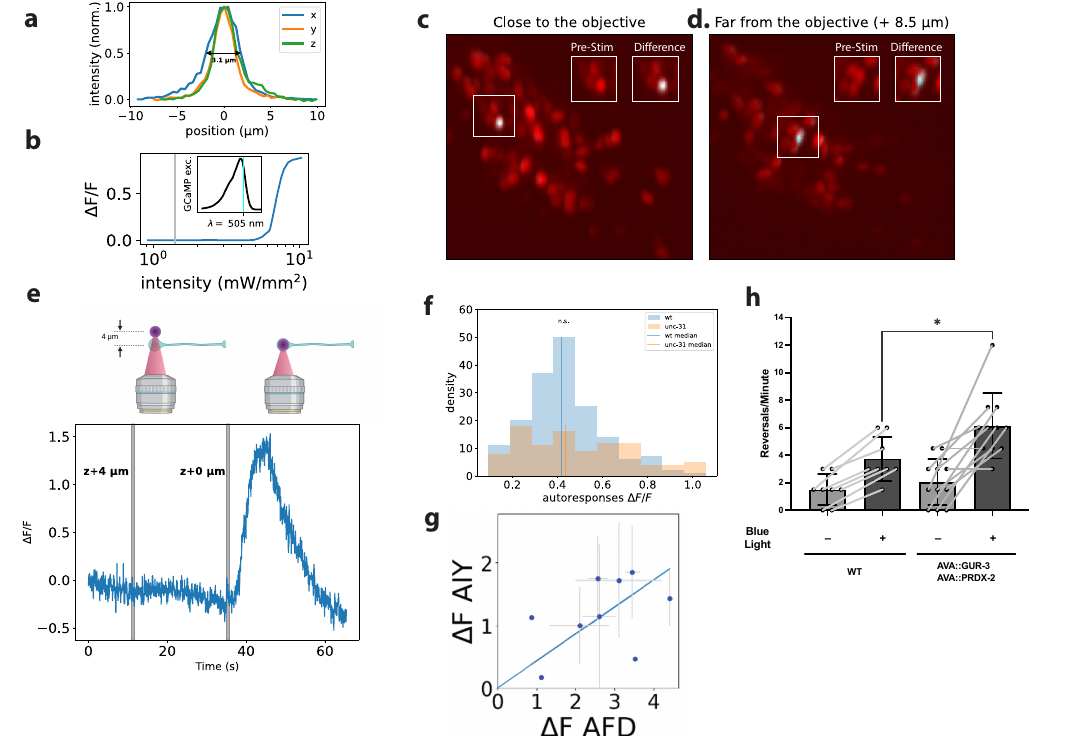}
\caption{ \textbf{Characterization of two photon optogenetic stimulation and evoked response.} \textbf{a)} Two-photon stimulation spot size.  \textbf {b)} The imaging excitation wavelength and intensity were chosen to  avoid GUR-3/PRDX-2 activation. GCaMP response to 500 nm activation of GUR-3/PRDX-2 expressing neuron as reported in \cite{bhatla_light_2015}. Vertical gray line indicates light intensity typically used for calcium imaging in the present work. Inset shows  GCaMP6 excitation spectra from \cite{chen_ultrasensitive_2013}.  Vertical cyan line indicates 505 nm excitation wavelength used in the present work. \textbf{c)} Selected neurons are  photobleached close to and \textbf{d)} far from the objective to demonstrate targeted illumination across the brain.  tagRFP-T is photobleached by 2p stimulation (20 s illumination, 200 $\mu$W at 500 kHz repetition rate, 3.1 $\mu$m diameter FWHM spot). Fluorescence difference image shows tagRFP-T expression merged with a false-color blue-green image to reveal the change in intensity before and after targeted illumination.    Only the targeted neuron and not nearby neurons appear photobleached. Insets shows zoomed-in image of the targeted neuron's original tagRFP-T intensity  (left) and difference image (right). Laser power  was chosen to avoid saturated bleaching.
\textbf{e)} \textit{In vivo} demonstration of the two-photon effective spot size.  Activity from a neuron expressing GUR-3/PRDX-2 and GCaMP6s  is shown in response to a 300 ms 2p stimulation delivered, at $t$=11s, 4 $\mu$m beyond the $\approx3.5$ $\mu$m diameter soma on the optical axis (z),  and at $t=35$s, centered on the soma ($t=35$s).
On-target soma stimulation, but not nearby stimulation, evokes a transient called an ``autoresponse.'' A stimulus artifact at $t=35$s is visible because no smoothing or filtering is applied to this trace. Schematic generated with BioRender.  \textbf{f)} Distribution of autoresponses under typical stimulation conditions, as described in methods (1.2 mW, 500 kHz; 0.5 s for WT, 0.3 s for \textit{unc-31}).  Only  stimulations that evoked activity meeting the criteria of an autoresponse are included, as described in methods. \textbf{g)} Measured calcium response of neuron AIY to optogenetic stimulation of AFD. Compare to Figure 4b in~\cite{narayan_transfer_2011}. A variety of stimulus durations was used to generate autoresponses of different amplitudes. \textbf{h)} Animals expressing GUR-3/PRDX-2 in AVA (n=11 animals) exhibited more blue-light evoked reversals per min than WT (n=8 animals). ~480 nm peaked light was delivered to freely moving animals. }  \label{sfig:spot_wavelength} \label{sfig:bleaching}\label{sfig:walkoff}\label{sfig:autoresponse_amplitude}\label{sfig:AFD-AIY}\label{ext:stim_characterization}
\end{figure}

\begin{figure}[htbp]
\centering
\includegraphics[page=3, width=1\linewidth]{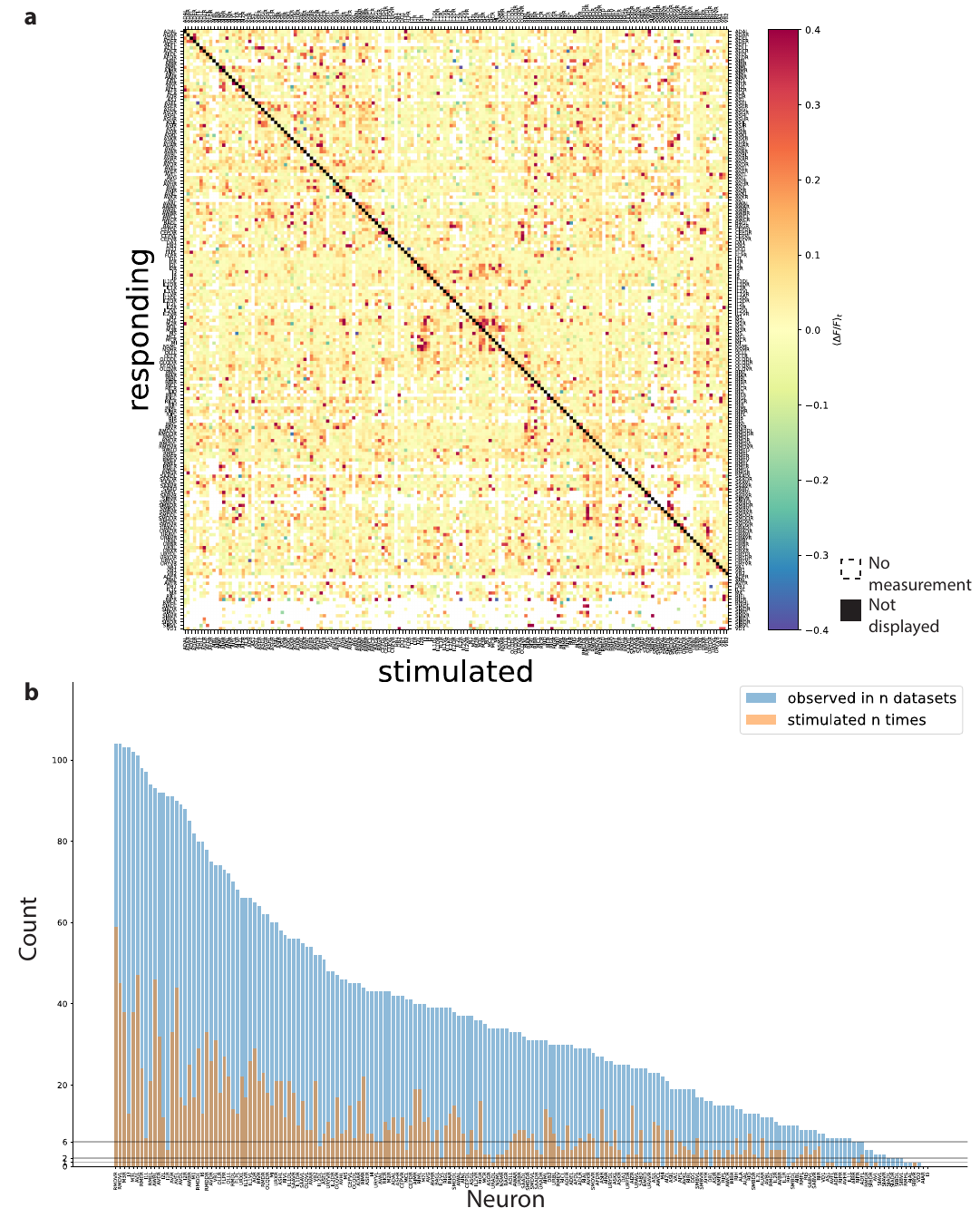}
\caption{\textbf{\Themap{}. a)}  Mean amplitude of neural activity in a post-stimulus time window ($\langle \Delta F/F_0 \rangle_t$) averaged across trials and individuals for WT background. White indicates no measurement.  For inclusion, a stimulus event is required to  evoke a response in the stimulated neuron. Therefore the strength of neuron's response to its own stimulation is not displayed (black diagonal). ($N$=\dataNumAnimals animals). \textbf{b}) Number of  stimulation events (orange) and number of datasets (animals) in which the neuron was observed (blue) for each neuron is shown. \label{sfig:num_obs_hist}}  \label{sfig:DF_only} \label{sfig:num_obs}
\end{figure}

\begin{figure}[htbp]
\centering
\includegraphics[page=4, width=.95\linewidth]{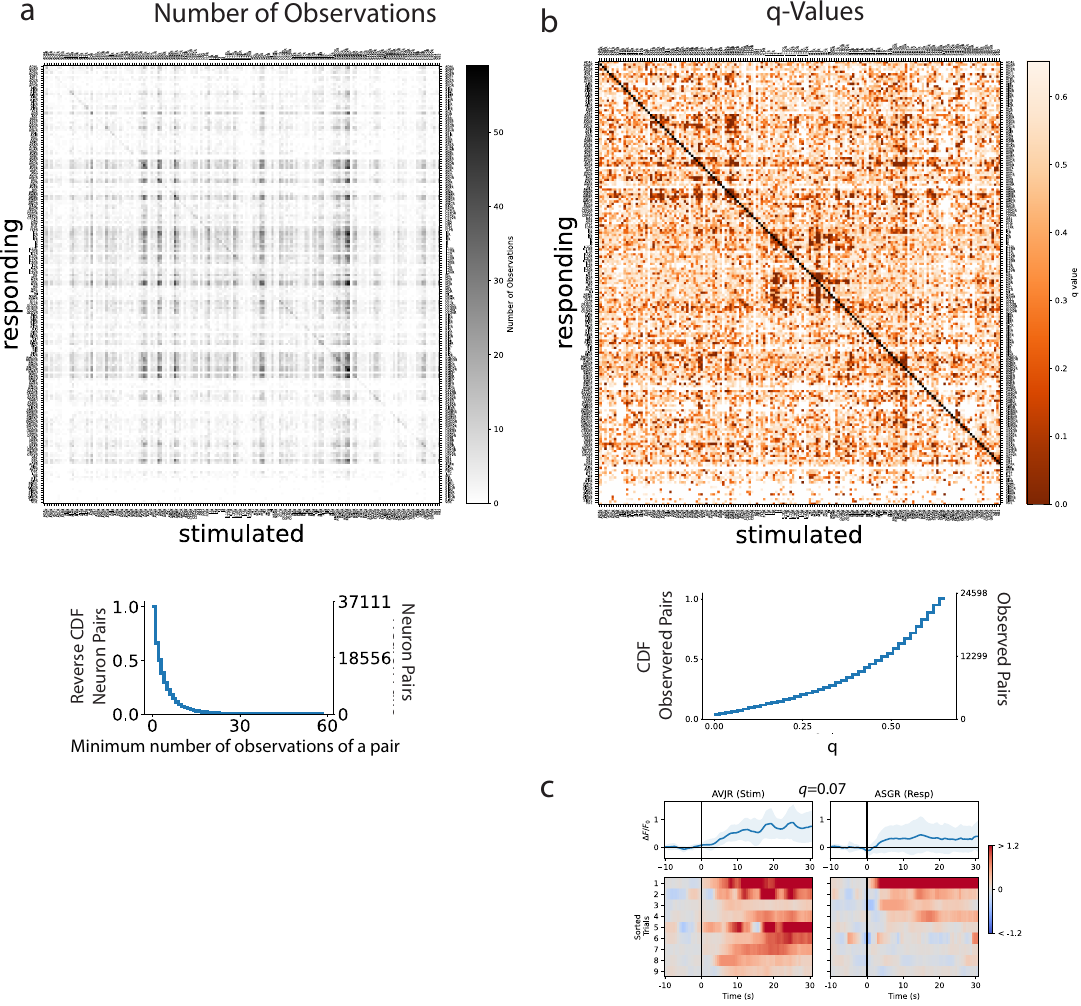}
\caption{ \textbf{Observations and false discovery rate of neuron pairs in the \themap{}.} \textbf{a)} Number of observations made of each neuron pair for WT-background animals. To be considered an observation, the upstream neuron must have been stimulated, calcium imaging of both the upstream and downstream neuron must have been recorded, and the upstream neuron must have exhibited an auto-response. Sorted as in Extended Data Figure~\ref{sfig:DF_only}a. Reverse cumulative distribution is also shown (bottom panel) and reports the  fraction of pairs (number of observed pairs divided by the total number of possible pairs of neurons in the head).\label{sfig:num_obs_matrix} \textbf{b)} $q$-values are shown for each neuron pair. $q$-values report false discovery rate of finding a functional connection. They provide a metric of significance for assessing  whether a neuron pair is functionally connected based on the number of observations and the magnitude of the response transients, taking into consideration the number of multiple hypothesis tested. Cumulative distribution is also shown (bottom panel). \textbf{c)} ASGR often exhibits activity immediately following stimulation of AVJR,  but because its $q$-value is greater than 0.05, it does not meet the stringent statistical threshold to be deemed ``functionally connected.'' }  \label{sfig:maps_wt} \label{sfig:q_only}\label{sfig:q}
\end{figure}

\begin{figure}[htbp]
\centering
\includegraphics[page=5, width=.8\linewidth]{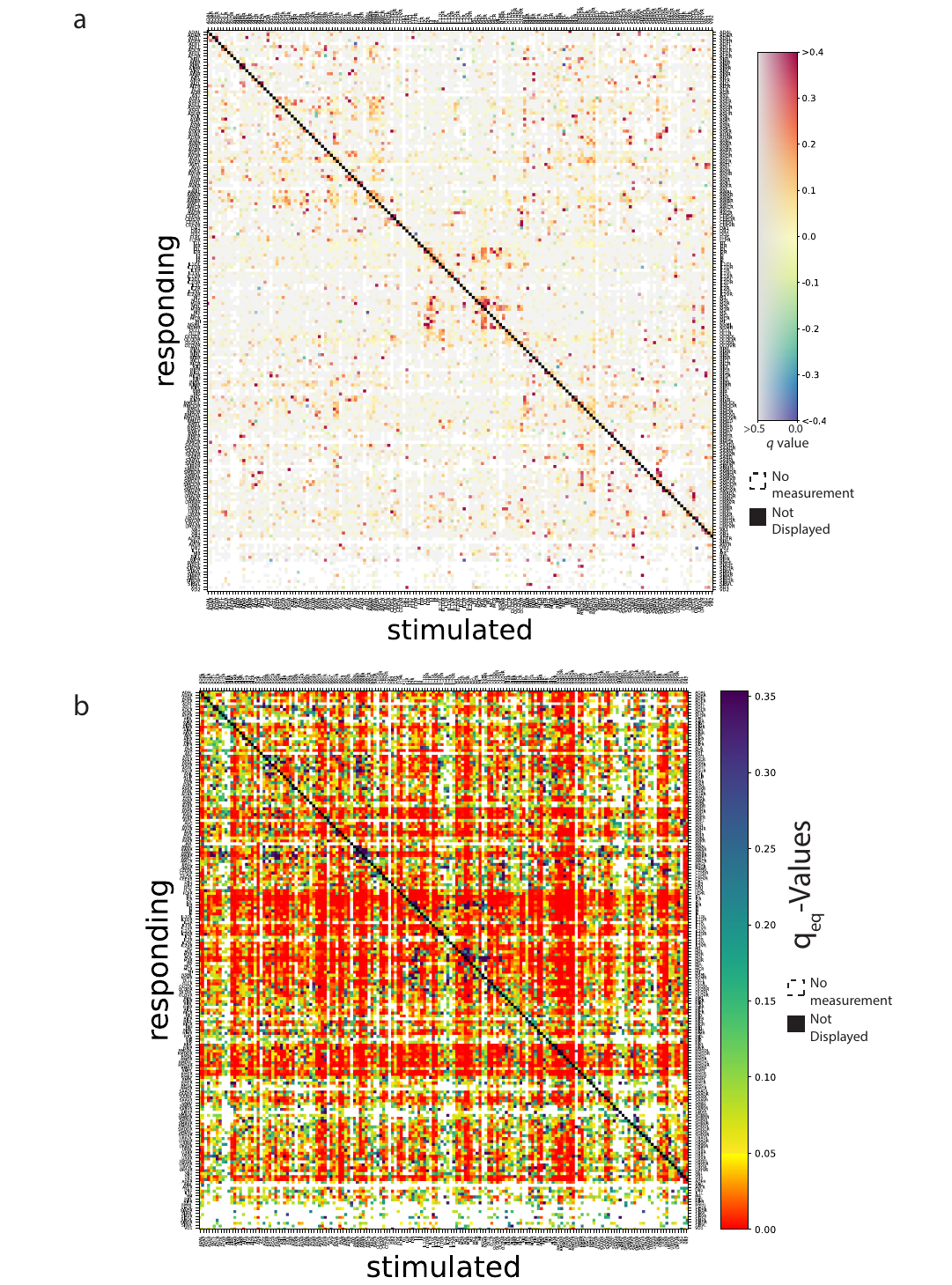}
\caption{ \textbf{ \Themap{}   showing false discovery rates for functional connections and non-connections. a)} Map of functional connections showing downstream calcium response amplitude and false discovery rate for WT.  Same as Fig.~\ref{fig:map} except here neurons that are  observed but not stimulated are also included.  Note the colorbar has two axes. Mean amplitude of   neural activity in a post-stimulus time window ($\langle \Delta F/F_0 \rangle_t$) averaged across trials and individuals is shown. $q$-value reports false discovery rate (more gray is less significant). White indicates no measurement.  Auto-response is required for inclusion and not displayed (black diagonal). ($N$=\dataNumAnimals animals). \textbf{b)} 
Map of functionally not connected pairs. The false discovery rate,  $q_\mathrm{eq}$,  is reported for declaring  a neuron pair to be \textit{not} functionally connected. Lower  $q_\mathrm{eq}$ (more red) indicates higher confidence that the observed downstream calcium activity is equivalent within a  bound $\epsilon$ to a null distribution of spontaneous activity. The false-discovery rate takes into consideration the amplitude of the calcium transient, the number of observations and the number of hypotheses tested. }  \label{sfig:q_eq} \label{sfig:maps_complete}
\end{figure}

\begin{figure}[htbp]
\centering
\includegraphics[page=6, width=.88\linewidth]{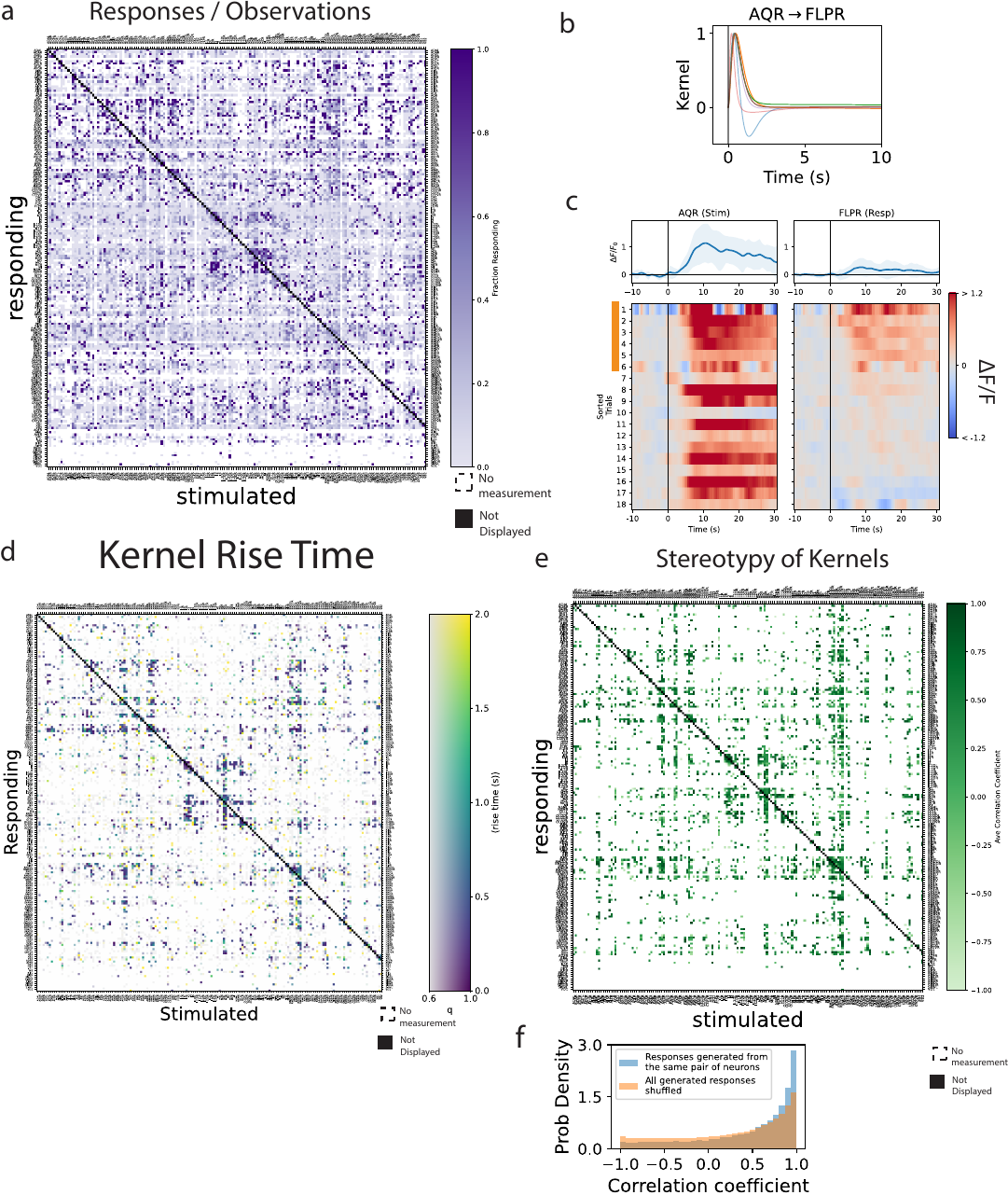}
\caption{ \textbf{ Timescales and variability of measured functional connectivity.} WT. \textbf{a)}  The fraction of  stimulation events that evoked a downstream ``response''  is shown for each neuron pair.  To be classified as a ``response'' requires a sufficiently large calcium transient amplitude and derivative. Auto-responses are required and not shown (black diagonal).   \textbf{b)} Kernels are functions that  return the downstream neuron's activity when convolved with the upstream neuron's activity. Kernels capture properties of the connection independent of  variability in the upstream neuron's auto-response. Kernels are shown for each \textbf{c)} FLP response to AQR stimulation.  Kernels are only calculated for stimuli that evoked downstream ``responses''  (indicated in orange).  \textbf{d)} Kernel rise time for each measured neuron pair in WT is a metric of  signal propagation speed. \textbf{e)} The stereotypy of kernels within each neuron pair is reported by calculating the average correlation coefficient among them. Only neuron pairs with at least two kernels are considered.
\textbf{f)}  Distribution of the correlation-coefficients of convolved kernels,  within each pair of neurons (blue, n=30,406), and across all kernels measured regardless of neuron pair (orange, n = 113,880,912).  }  \label{sfig:variability}\label{sfig:kernels}
\end{figure}

\begin{figure}[htbp]
\centering
\includegraphics[page=7, width=\linewidth]{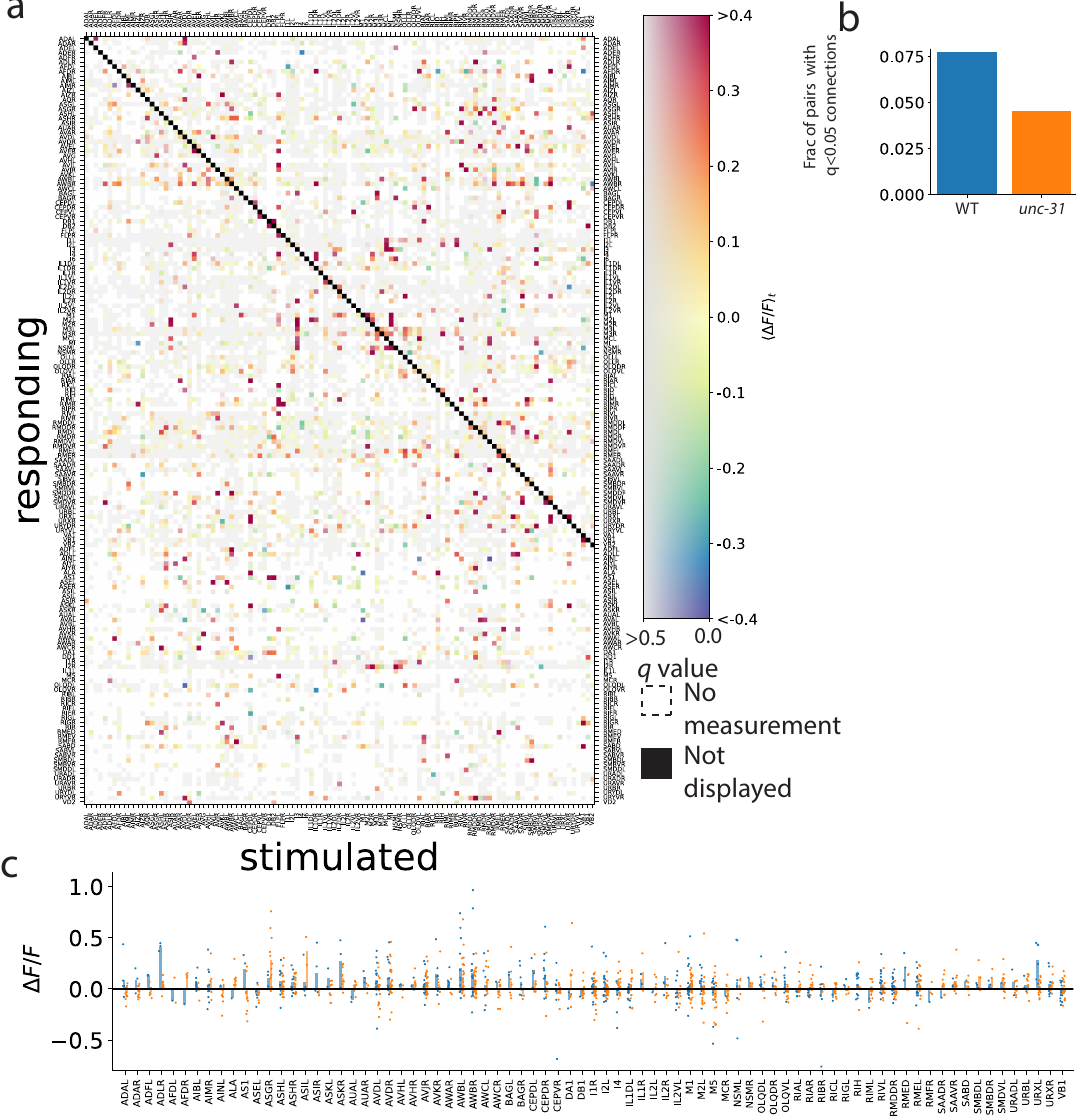}
\caption{\textbf{Signal propagation of \textit{unc-31} background, with defects in dense-core vesicle-mediated extrasynaptic signaling.} \textbf {a)} Same format as Extended Data Figure~\ref{sfig:maps_complete}a. Mean amplitude of   neural activity in a post-stimulus time window ($\langle \Delta F/F_0 \rangle_t$) averaged across trials and individuals is shown. $q$-value reports false discovery rate and is a metric of significance (more gray is less significant). White indicates no measurement.  Auto-response is required and not displayed (black diagonal).  ($N$=18 animals). \textbf{b)} \textit{unc-31} mutants  had a smaller proportion of measured pairwise neurons that were functionally connected ($q<0.05$) than WT (comparison run on pairs for which data is present in both WT and \textit{unc-31} mutants).
\textbf{c)} Responses to RID stimulation are shown for WT (blue) and \textit{unc-31} (orange). Points are responses, bar is mean across trials and animals. Neurons with the smallest amplitude responses are not shown.   Corresponding traces for  ADLR, AWBR and URXL are shown in Fig.~\ref{fig:wireless}. As in that figure, responses here are shown even for those cases when RID's calcium activity was not measured and therefore do not appear in a. Different inclusion criteria are used here to accommodate cases where the tagRFP-T expression is dim, as described in methods.}  \label{sfig:maps_unc31}\label{sfig:unc31_vs_WT_stats}\label{sfig:RID_more_responses}
\end{figure}

\begin{figure}[htbp]
\centering
\includegraphics[page=8, width=.95\linewidth]{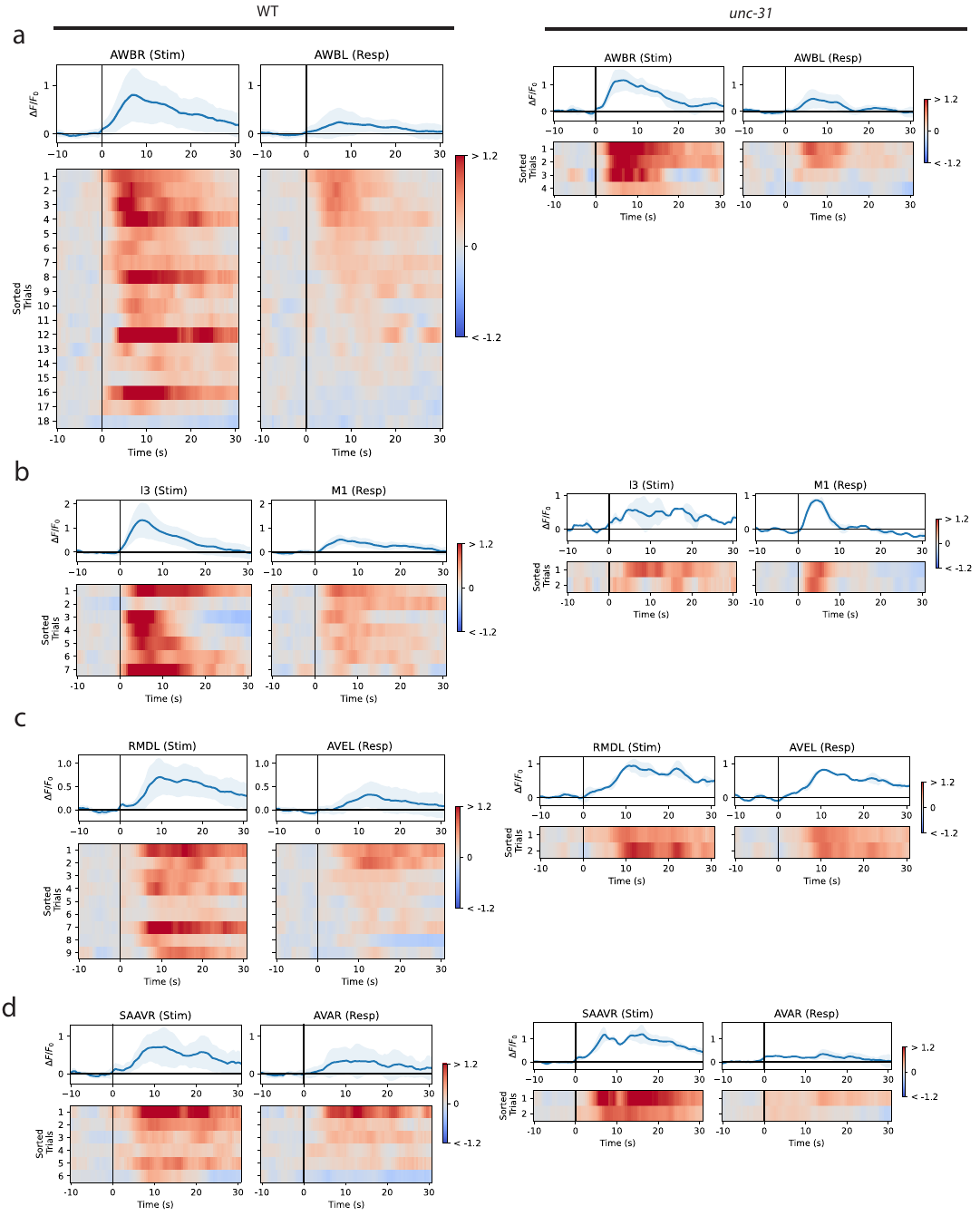}
\caption{\textbf{ Neural responses for some pairs are similar in wild-type and \textit{unc-31} mutant animals.} Paired stimulus and response  traces of selected neuron pairs with \textbf{a-c)} monosynaptic  gap junctions or \textbf{d)} monosynaptic chemical synapses are shown in WT background (left) and \textit{unc-31} mutant background (right).}  \label{sfig:unc31_vs_WT_control}  
\end{figure}

\begin{figure}[htbp]
    \centering
    \includegraphics[page=9,width=\linewidth]{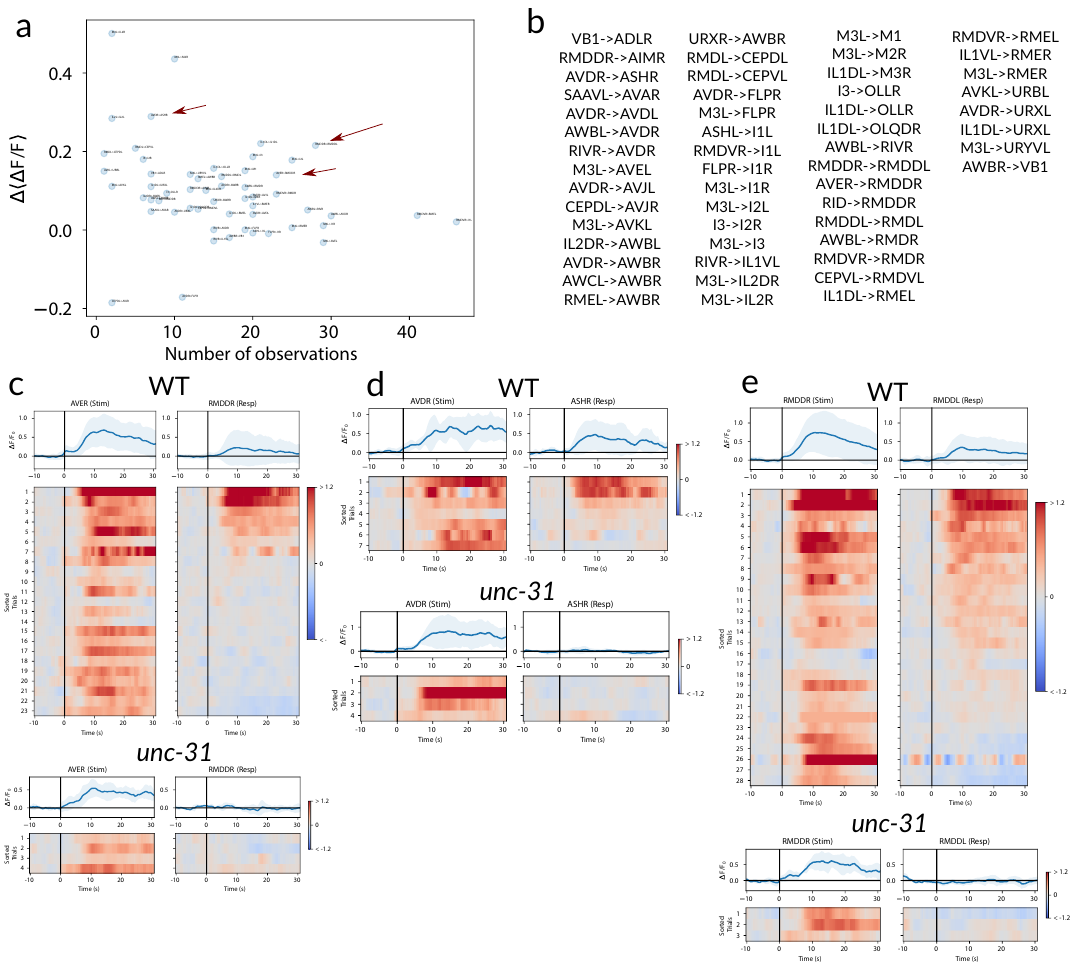}
    \caption{
    \textbf{Examples of candidate purely extrasynaptic pairs. a)}$\langle\Delta F/F\rangle$ vs number of observations for our candidate purely extrasynaptic-dependent pairs. Arrows indicate examples shown below. \textbf{b)}List of candidate entirely extrasynaptic-dependent connections. Relevant neuropeptide  GPCR expression is  listed in Supporting Spreadsheet 1, compiled from~\cite{beets_system-wide_2022,taylor_molecular_2021}, following~\cite{ripoll-sanchez_neuropeptidergic_2022}.
    \textbf{c-e)} Paired responses in WT and \textit{unc-31} animals for the candidate extrasynaptic pairs AVER->RMDDR, AVDR->ASHR, and RMDDR->RMDDL respectively, selected among all the candidates as illustrated in panel (a)}
    \label{sfig:esyn}  \label{stab:esyn}
\end{figure}

\begin{figure}[htbp]
    \centering
    \includegraphics[page=10,width=\linewidth]{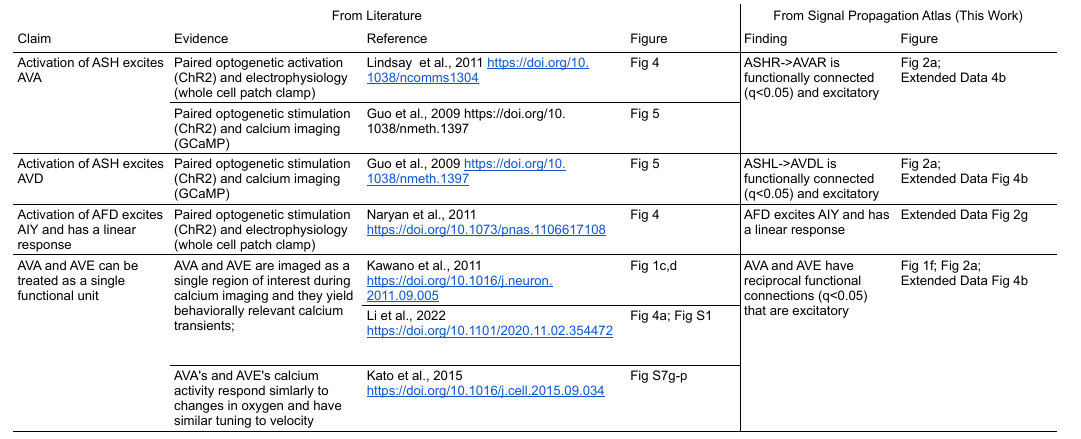}
    \caption{
    \textbf{Selected instances of agreement between measured signal propagation and previously reported functional measurements.}}
    \label{ed:agreement_table}  
\end{figure}

\clearpage

\section*{Supplementary Material}
\paragraph{Supplementary Spreadsheet 1}\url{https://doi.org/10.6084/m9.figshare.23057558} Microsoft Excel file lists neuropeptide and GPCR combinations for putative purely extrasynaptic-dependent signaling pairs of neurons.    Neuropeptide and GPCR combinations are automatically generated based on data from  \cite{beets_system-wide_2022,taylor_molecular_2021}, following~\cite{ripoll-sanchez_neuropeptidergic_2022}.]

\paragraph{Supplementary Movie 1} M1 stimulation evokes pharyngeal muscle contraction.   Video from four animals are shown during signal propagation measurements. In each example, neuron M1 is optogenetically stimulated for 0.5 s and  pharyngeal muscle contractions are visible by their effect on the location of neighboring neurons.  M1 is known to release acetylcholine via chemical synapse onto pharyngeal muscles  \cite{franks_comparison_2009, sando_hourglass_2021}. Top panel shows z-projection of tagRFP-T. Bottom panel shows z-projection of GCaMP6 fluorescence. Crosshairs indicate neuron currently targeted for stimulation.
\end{document}